\documentstyle[12pt]{article}
\begin{document}
\newcommand{\ket}[1] {\mbox{$ \vert #1 \rangle $}}
\newcommand{\bra}[1] {\mbox{$ \langle #1 \vert $}}
\newcommand{\bkn}[1] {\mbox{$ < #1 > $}}
\newcommand{\bk}[1] {\mbox{$ \langle #1 \rangle $}}
\newcommand{\scal}[2]{\mbox{$ < #1 \vert #2 > $}}
\newcommand{\expect}[3] {\mbox{$ \bra{#1} #2 \ket{#3} $}}
\newcommand{\ki}{\mbox{$ \ket{\psi_i} $}}
\newcommand{\bi}{\mbox{$ \bra{\psi_i} $}}
\newcommand{\p} \prime
\newcommand{\e} \epsilon
\newcommand{\la} \lambda
\newcommand{\om} \omega   \newcommand{\Om} \Omega
\newcommand{\cc}{\mbox{$\cal C $}}
\newcommand{\w} {\hbox{ weak }}
\newcommand{\al} \alpha
\newcommand{\bt} \beta

\newcommand{\be} {\begin{equation}}
\newcommand{\ee} {\end{equation}}
\newcommand{\ba} {\begin{eqnarray}}
\newcommand{\ea} {\end{eqnarray}}

\def\lrD{\mathrel{{\cal D}\kern-1.em\raise1.75ex\hbox{$\leftrightarrow$}}}
\def\lr #1{\mathrel{#1\kern-1.25em\raise1.75ex\hbox{$\leftrightarrow$}}}

\overfullrule=0pt \def\sqr#1#2{{\vcenter{\vbox{\hrule height.#2pt
          \hbox{\vrule width.#2pt height#1pt \kern#1pt
           \vrule width.#2pt}
           \hrule height.#2pt}}}}
\def\square{\mathchoice\sqr68\sqr68\sqr{4.2}6\sqr{3}6} \def\lrpartial{\mathrel
{\partial\kern-.75em\raise1.75ex\hbox{$\leftrightarrow$}}}

\begin{flushright}
LPTENS 95/34\\
September 1995
\end{flushright}
\vskip 1. truecm
\vskip 1. truecm
\centerline{\LARGE\bf{The Recoils of a Dynamical Mirror}}
\vskip 2 truemm
\centerline{\LARGE\bf{and the Decoherence of its Fluxes}}
\vskip 1. truecm
\vskip 1. truecm

\centerline{{\bf R. Parentani}\footnote{e-mail: parenta@physique.ens.fr}}

\vskip 5 truemm
\centerline{Laboratoire de Physique Th\'eorique de l'Ecole
Normale Sup\'erieure\footnote{Unit\'e propre de recherche du C.N.R.S.
associ\'ee \`a l'
ENS et \`a l'Universit\'e de Paris Sud.}}
\centerline{24 rue Lhomond
75.231 Paris CEDEX 05, France.}
\vskip 5 truemm
\vskip 1.5 truecm

\vskip 1.5 truecm
\vskip 1.5 truecm
\vskip 1.5 truecm
{\bf Abstract }
In order to address the problem of the validity of the "background
field approximation",
we introduce a dynamical model for a mirror described by a massive
quantum field. We then
analyze the properties of the scattering of a massless field
from this dynamical mirror and
compare the
results with the corresponding quantities evaluated using the original Davies
Fulling
model in which the mirror is represented by
a boundary condition imposed on the massless field at its surface.
We show that in certain circumstances, the recoils of the dynamical mirror
induce decoherence effects which subsist even when the mass of the mirror is
sent to infinity.
In particular we study the case of the uniformly accelerated mirror and prove
 that, after a certain lapse of proper time,
 the decoherence effects inevitably dominate
 the physics of the quanta emitted forward. Then,
the vanishing of the mean flux obtained in the Davies Fulling model is no
longer found
but replaced by a positive incoherent flux.
\vfill \newpage

\section{Introduction}
In deriving quantum black hole radiance, Hawking\cite{Hawk} realized from the
outset that the
frequencies defined on $\cal I^-$
and involved in the Bogoliubov coefficients relating out-modes
to in-modes grow exponentially fast with the retarded time around which the
out-particle is created.
 Accordingly Hawking
justifies the use of "geometrical optics"
to describe propagation through
the star. For a recent presentation of this point of view see ref. \cite{new}.

More recently there have been
debates\cite{THooft}\cite{Jacobson}\cite{Waldbook}
 on the relevance of these
"transplanckian" frequencies when one wishes to take into account
gravitational backreaction effects beyond the semi-classical theory
wherein only the mean value of the energy momentum tensor acts as a source in
Einstein's
equations\cite{Bardeen}\cite{York}\cite{Massar2}\cite{PP}.
 We remind the reader that the quantum averaged
stress tensor is perfectly regular\cite{DFU}
 in the evaporating geometry (until the residual black hole mass approaches
the Planck mass) and that an infalling observer
would detect no quantum characterized by these high frequencies as it crosses
the horizon\cite{Hawk}.

Whatever is the reader's opinion, the fact is that
when ones uses the two assumptions made by Hawking, to wit {\it  free}
propagation
in a {\it given} geometry, one finds
 quantum correlations between the detection of an asymptotic
quantum and the configurations at early times which are characterized by these
frequencies. These correlations show up in various ways. As said above, they
appear in the Bogoliubov coefficients. It is nevertheless not straightforward
to
understand the meaning nor the consequences of this presence. More explicit is
the calculation of
the commutators between local operators and asymptotic annihilation operator
as proposed in ref. \cite{THooft2} or equivalently the calculation of
off-diagonal matrix elements
of $T_{\mu \nu}$ as done in refs. \cite{EMP}\cite{MaPa}\cite{Verl3}.
Indeed, in a perturbative treatment of the gravitational back-reaction, these
matrix
elements do intervene in the first order corrections.

The challenging problem is therefore to determine when the gravitational
effects
induced by these fluctuations and
absent in the semi-classical treatment will invalidate the predictions of
the latter. The problem is not to ask {\it in abstracto} whether the background
field approximation
is valid or not. Rather, one should work with different schemes of calculation
to approach the fully quantized theory.
Then one tests the validity of the semi-classical treatment
by determining the duration of time for which it furnishes a correct estimate
of the matrix elements
of the relevant observables.
%

The goal of this paper is to answer this type of questions in the simpler
context
of the radiation spontaneously emitted by a non inertial mirror.
Two different schemes will be compared. First, we shall reexamine the original
Davies and Fulling
(DF) model\cite{DF} in which the trajectory of the mirror is classical and
specified once for all.
This corresponds to the background field approximation without backreaction.
Secondly, we shall introduce a quantized version
in which the mirror is
described by a massive scalar quantum field; see ref. \cite{rec}
for a similar quantization in the case of the accelerated detector.
In this second scheme, the mirror recoils according to Feyman rules when it
scatters a quantum or
when it creates a pair of quanta. In this we differ from Chung and
Verlinde\cite{CV}
who work with a hybrid (semi-classical) model
in which the source of the recoil is the mean energy momentum tensor expressed
in terms of the (first) quantized mirror's trajectory. This point is further
taken up in Section 5.
We then
 determine from our formalism
what is the subset of physical quantities correctly estimated in the DF model
in various circumstances. It will be seen that
the matrix elements which are
 sensitive to interferences (i.e. sensitive to relative phases between states
characterized by different particle number) will be the first affected by
the recoils of the heavy mirror. On the contrary, those expressed as sum
of squares will be almost unaffected by the recoils and therefore almost
identical to the ones
obtained in the DF model.

We have organized the paper as follows.
In Section 2 we review the properties of the DF model emphasizing the
fluctuating
properties of the flux spontaneously emitted.

In Section 3, we analyze the flux emitted by a uniformly accelerated mirror. It
was shown by
Davies and Fulling that the mean flux vanishes even though
there are emissions of quanta, see \cite{BD}\cite{Grov}. We show in detail how
these antagonistic
properties can be understood by introducing the concept of a partially
reflecting mirror
which allows us to decouple adiabatically the field from the mirror for
asymptotic times.
Furthermore, we relate these properties to
Unruh effect\cite{Unr} and to the radiation emitted by a thermalized
accelerated
atom which enjoys similar
properties\cite{UnrW}\cite{grove}\cite{Unruh92}\cite{AM}\cite{MaPa}.

In Section 4, we introduce our model for an inertial dynamical mirror
based on quantum field theory. Since the mirror is inertial there is no
production but the scattering
is however not trivial. We show how this dynamical model is related to the
DF model by taking the large mass limit of the $S$ matrix
elements and then resumming the Born series\cite{mt} so as to
 recover the overlap between in and out
 states evaluated in the DF model. Nevertheless, in spite of this strict
correspondence, interferences effects between the radiation and the mirror wave
functions
 are induced by recoil
 (i.e. the momentum transfer upon reflection) and modify the expectation values
of local operators
when the radiation is characterized by a fluctuating particle number.
 Would the radiation be described by a diagonal matrix density or by a
pure state with a definite particle number, no such effect would be found
 in the large mass limit.

In Section 5, we generalize the dynamical model to non-inertial trajectories by
introducing an
external electric field which brings a charged mirror into non-inertial motion.
We reexamine
the case of constant acceleration and compare the results with the ones
obtained in Section 3.
We see how the recoils induced by the production acts modify the properties of
the
flux.
The main consequence is that the vanishing of the mean flux can no longer be
maintained for an arbitrarily long period. After a proper time lapse given by
$(\ln M/a)/2a$,
where $a$ is the acceleration and $M$ the rest mass of the mirror, the quanta
emitted forward
decohere and the flux of energy is positive.
Furthermore, the notion of the "partner" of a specific quantum (particles are
always created in pairs)
is also affected by the dynamical character of the mirror. Indeed, the mirror
gets correlated to the
constituents of each pair by the momentum transfer at the creation act.
To describe these correlations we consider the conditional value\cite{MaPa}
 of the energy momentum tensor correlated to the detection of a specific
quantum. This quantity
which is expressed as an off-diagonal matrix element of $T_{\mu \nu}$
is much more sensitive  to the dynamics of the mirror than the mean value.
It is therefore a useful tool to investigate the role of the dynamics of the
mirror
and as a byproduct to understand the nature and the validity of the "background
field
approximation" represented here by the DF model.

\section{The Scattering Amplitudes and the Fluxes in the Absence of Recoil}

In this section, we review the properties of the Davies-Fulling model\cite{DF}
wherein a massless quantum field is scattered by a moving
mirror which follows a given classical trajectory.
We emphasize the relations between the classical and the quantum
aspects of this scattering because they illustrate the nature of the
approximations implicitly used in the DF model when one considers it
as the large mass limit of some dynamical model.
For the same reason, we also describe the fluctuating
properties of the flux spontaneously emitted by the mirror.
The reader may consult refs. \cite{DF}\cite{BD}\cite{Grov} for discussions
of others aspects of the DF model.

Following Davies and Fulling \cite{DF},
we consider the quantum mechanical problem defined
by the two equations
\be
\left[ \partial_t^2 -\partial_z^2 \right] \phi(t,z) =0
\label{one}
\ee
and
\be
\phi(t, z_{cl}(t)) =0
\label{two}
\ee
where $\phi $ is a complex field and $z_{cl}(t)$ is the trajectory
of the "mirror" expressed in carthesian coordinates $t$ and $z$. Since we work
in
$1+1$ dimensions, it is appropriate to introduce the light-like coordinates
$U$ and $V$ defined by
$U, V = t \mp z$. In these coordinates
 eq. (\ref{one}) becomes $\partial_U \partial_V \phi(U,V) =0$, hence the
general solution is
the sum of functions of $U$ or $V$ only.

Owing to the linearity of the both eqs. (\ref{one}) and (\ref{two}) {\it
and} the classical character of the given trajectory,
the quantum scattering amplitudes can be obtained in purely classical terms in
the sense
that $\hbar$ does not appear.
We shall therefore
first analyze the classical theory of the scattering of $\phi$ and then
see how the classical concepts are reinterpreted in second quantization.
This exercise will be found useful when, in the next sections, we
shall modify eqs. (\ref{one}) and (\ref{two}) to give a dynamical content to
the mirror's motion.
In particular it will allow us to identify the corrections which are quantum.

In this section, for simplicity, the mirror's trajectory is taken to be
inertial and time-like at $t=\pm \infty$. In conformal terms, it means that it
starts in $i^-$ and ends up in $i^+$\cite{MTW}, the time-like past and future
infinities.
In scattering terms, it means that there is a complete decoupling of what
happens
to the right and to the left. Therefore the current of an ingoing
flux is fully recovered in the scattered outgoing flux
and {\it vice versa} if one studies propagation backwards in time.
When the trajectory does not end up at $i^+$, there is, so to speak, the
formation of a horizon and a part only of the incident flux is
reflected. This situation will be discussed in the next section.

We first study the scattering to the right of a left moving
in-mode of frequency $\om$
\be
\varphi_{\om}(V) ={ e^{-i \om V} \over \sqrt{4 \pi \vert \om \vert } }
\label{three}
\ee
This mode carries a conserved current $J_V= \varphi_{\om}^* i\! \lr{\partial_V}
\varphi_{\om}= 1/2 \pi$ and its norm is given by
\be
\bkn{\varphi_{\om^\p},\ \varphi_{\om}}=\int^{\infty}_{-\infty} dV
\varphi_{\om^\p}^*i\!\lr{\partial_V}
\varphi_{\om}= \mbox{sign} (\om) \delta (\om^\p - \om)
\label{four}
\ee

Upon reflection on the mirror whose trajectory is now expressed as
$V=V_{cl}(U)$, the scattered mode is
\be
\varphi_{\om}^{scat}(U)= -
{ e^{-i \om V_{cl}(U)} \over \sqrt{4 \pi \vert \om \vert } }
\label{five}
\ee

On ${\cal I^+}$, i.e. on $V=\infty$, one can decompose this scattered mode
in Fourrier transform in terms of the right moving out-modes
\be
\varphi_{\la}(U) = { e^{-i \la U} \over \sqrt{ 4 \pi \vert \la \vert }}
\label{six}
\ee
 which form a complete set if $\la $ span the whole real axe:
\be
\varphi^{scat}_{\om}(U) = \int^{\infty}_{-\infty} d\la
\gamma_{\om, \la}
\varphi_{\la}(U)
\label{seven}
\ee
where the coefficents $\gamma_{\om, \la} $ are given by
\ba
\gamma_{\om, \la} &=& \mbox{sign}(\la)
\bkn{ \varphi_{\la},\ \varphi_{\om}}\nonumber\\
&=& - 2 \int^{\infty}_{-\infty} dU { e^{i \la U} \over \sqrt{ 4 \pi \vert \la
\vert^{-1} }}
{e^{-i \om V_{cl}(U)} \over \sqrt{ 4 \pi \vert \om \vert }}
\label{eight}
\ea
We have used eq. (\ref{four}) and integrated by parts.
Eqs. (\ref{seven}) and (\ref{eight}) express the conventional Fourrier
decomposition with the relativistic weight $1/\sqrt{4 \pi \vert \la \vert}$
included. This writting together with the conservation of the
current on the mirror implies that the coefficients $\gamma_{\om, \la}$
satisfy the following "unitary" relations
\begin{eqnarray}
\int_0^\infty d\la \left[ \gamma_{\om, \la} \gamma^*_{\om^\p, \la} -
\gamma_{\om, -\la}\gamma^*_{\om^\p, -\la}\right]
 &=& \delta (\om^\p -\om) \nonumber\\
\int_0^\infty d\om \left[ \gamma_{\om, \la^\p}\gamma^*_{\om, \la} -
\gamma^*_{-\om, \la^\p}\gamma_{-\om, \la}\right]
 &=& \delta (\la^\p - \la)
\label{nine}
\end{eqnarray}

In classical mechanics, the energy density of an incident wave of
frequency $\om $ and amplitude $A_{\om}$ is
\be
T_{VV}^{cl} = 2 \vert A_{\om}\vert^2 \ \vert \partial_V \varphi_{\om}(V)
\vert^2 =
\vert A_{\om}\vert^2
{\vert \om \vert \ \over 2 \pi }
\label{Tvv}
\ee
Using eq. (\ref{seven}), the energy density of the scattered mode reads
\begin{eqnarray}
T_{UU}^{cl} &=& 2
\vert A_{\om}\vert^2 \ \vert \partial_U \varphi_{\om}^{scat}(U) \vert^2
\nonumber\\
&=&
\vert A_{\om}\vert^2
 \int_0^\infty d\la \int_0^\infty d\la^\p {\sqrt{ \la \la^\p} \over 2\pi}
\nonumber\\
&&\quad\quad\quad\left[
e^{-i(\la -\la^\p) U} \left(\gamma_{\om, \la}
 \gamma_{\om, \la^\p}^*
+ \gamma_{\om, -\la}^* \gamma_{\om, -\la^\p} \right)
- 2 \mbox{Re}
\left( e^{-i(\la + \la^\p) U} \gamma_{\om, \la}
\gamma_{\om, -\la^\p}^* \right) \right]
\label{Tuu}
\ea
Notice that the second term does not contribute to the total scattered
energy defined by $H_U
= \int^\infty_{-\infty} dU T_{UU}^{cl}$.

In order to get an approximate temporal description, we
 introduce broad wave packets defined by
\be
\varphi_{\bar \om}(V) = \int_{-\infty}^{\infty} d\om^\p f_{\bar \om}(\om^\p)
\varphi_{\om^\p} (V)
\label{broad}
\ee
These waves carry unit charge if $\int d\om \vert f_{\bar \om} (\om^\p) \vert^2
=1$.
They are broad if the spread around the mean frequency $\bar \om$ is
much smaller than $\bar \om$ itself.
Broad wave packets of the form $A_{\bar \om} \varphi_{\bar \om}$
carry an energy given by
$H_{V}= \int  dV T_{VV}^{cl} = \vert A_{\bar \om} \vert^2 \vert \bar \om \vert
$.

The classical relation
between the Doppler shifted frequency
of the scattered mode and the out-frequency $\la$ is recovered in this
formalism upon computing the stationary phase in the integrant of eq.
(\ref{seven}). One obtains
\be
\om {dV_{cl} (U) \over dU} = \la
\label{ten}
\ee
At fixed $\la$ and $\om$, it indicates at which value of $U$ does
$\om$ resonates with $\la$. (Remember that $dV_{cl}/dU >0$ holds for time like
trajectories.)
Therefore, when the rate of change of the trajectory  is small on the
scale of the inverse frequency of the incoming wave packet, i.e. when the
adiabatic (W.K.B.)
approximation is valid, the mean frequency of the
scattered mode
issuing from a broad wave packet
centered along $V=V_0$ and of mean frequency $\bar \om$ is
given by $\bar \la = \bar \om dV_{cl}/dU$ evaluated on
$V=V_0$. The mirror acts therefore as an infinite reservoir of energy
and momentum which leads to these relations no matter what $A_\om$ and $\om$
are.
Upon introducing some dynamics, we shall see that the rest mass
of the mirror will modify these relations for incoming waves with large
energy, i.e. comparable to the mirror's mass.

Before studying the quantum scattering, we notice that, for broad
wave packets, the charge
carried by the first term of eq. (\ref{nine}) is greather than the incident
one.
A similar increase of the reflected wave found in a Kerr geometry for certain
angular momenta\cite{ker}
was called super-radiance by analogy with induced emission.

\vskip 1. truecm

{\it Quantum Scattering}

\noindent
In second quantization, $\phi$ becomes a field operator.
In the Heisenberg representation, it is decomposed in positive and negative
frequency parts according to
\be
\phi(U,V) = \int_0^\infty d\om \left( a_\om \left[ \varphi_{\om} (V) +
\varphi_{\om}^{scat}(U) \right]
+b^\dagger_\om \left[ \varphi^*_{\om} (V) + \varphi^{*\ scat}_{\om}(U) \right]
\right)
\label{in}
\ee
The annihilation operators $a_\om, b_\om$ define the in-vacuum
 by $a_\om \ket{0, in} =b_\om \ket{0, in}=0$
\cite{DF}\cite{BD}.
This in vacuum corresponds to the usual Minkowski vacuum on ${\cal I^-}$ since
the
trajectory of the mirror is asymptotically inertial.

Similarly one decomposes $\phi$ in the out-basis
\be
\phi(U,V) = \int_0^\infty d\la
\left( a_\la \left[ \varphi_{\la} (U) + \varphi_{\la}^{bscat}(V) \right]
+ b^\dagger_\la \left[ \varphi^*_{\la} (U) + \varphi^{*\ bscat}_{\la}(V)
\right]
\right)
\label{out}
\ee
where the $a_\la$ and the $b_\la$ operators define the out-vacuum and where
$\varphi_{\la}^{bscat}(V)$ is the back scattered wave obtained
by propagating backwards the out-mode $\varphi_{\la} (U) $ and
imposing the reflection condition, eq. (\ref{two}).
It is given by
\be
\varphi_{\la}^{bscat}(V) = - {e^{-i\la U_{cl}(V)} \over \sqrt{4 \pi \vert \la
\vert }}
\label{bscat}
\ee

The $a_\la$ and the $b_\la$ operators define the out-vacuum which
correspond to Minkowski vacuum on $\cal{I^+}$.
The Bogoliubov transformation relates the in operators to the out ones as
\begin{eqnarray}
a_{\om} &=& \al^*_{\om, \la} a_\la + \bt^*_{\om, \la} b_\la^\dagger\nonumber\\
b_{\om} &=& \al^*_{\om, \la} b_\la + \bt^*_{\om, \la} a_\la^\dagger
\label{bog}
\end{eqnarray}
where the Bogoliubov coefficients $\al_{\om, \la}$ and $\bt_{\om, \la}$ are
related to
the $\gamma_{\om, \la}$ by
\begin{eqnarray}
\al_{\om, \la}&=& \gamma_{\om, \la}\nonumber\\
\bt_{\om, \la} &=& \gamma_{\om, -\la}
\label{albega}
\end{eqnarray}
Therefore the values of $\al$ and $\bt$ have nothing to do with quantum
mechanics, in particular $\bt_{\om, \la}$ where $\om $ and $\la$ are
 frequencies
and not energies does not become comparable to $\al_{\om, \la}$ as $\hbar \to
0$.

In quantum mechanics,
the non vanishing character of the $\bt$ coefficients encodes
spontaneous pair creation. Indeed, starting
with no particle on ${\cal I^-}$, the probability to
find no particle on ${\cal I^+} $ is no longer unity
(as it would have been if the trajectory was inertial)
but given by
\be
\vert \bk{0, out \vert 0, in} \vert^2
= e^{-\Sigma_{\bar \la} \ln (1 + \bk n_{\bar \la})}
=\Pi_{\bar \la} { 1\over \int d\om \vert \al_{\om, \bar \la} \vert^2 }
={ 1 \over Z } <1
\label{inout}
\ee
where $\bk n_{\bar \la}$ is the
 mean number of created particle of frequency $\bar \la$. In terms of the
Bogoliubov coefficients it is given by
\be
\langle n_{\bar \la} \rangle =
\bra{0, in} a^\dagger_{\bar \la} a_{\bar \la} \ket{0, in} = \int_0^\infty d\om
\vert  \bt_{\om, {\bar \la}}\vert ^2
 \ (=\bra{0, in} b^\dagger_{\bar \la} b_{\bar \la} \ket{0, in} )
\label{meannum}
\ee
We have used for conveniance of complete set of broad wave packets
labelled by $\bar \la$ to represent the out states. Their usual normalization
in terms of $\delta$
of Kronecker simplify both the equations and their interpreation.

Returning to plane waves,
one defines the mean density
by $\bk{n_\la} d\la = \bk{ n_{\bar \la}}$. Then, in terms of this density,
the mean energy emitted by the mirror is given by
\be
\bk {H_U^{scat} }= 2 \int_0^\infty d\la \la
\langle n_{\la} \rangle =
 \int_0^\infty d\la \la \bra{0, in} \left( a_\la^\dagger a_\la +
b_\la^\dagger b_\la \right) \ket{0, in}
=\int_{-\infty}^\infty dU \bk{ T_{UU}^{scat}}
\label{meanen}
\ee
where $ \bk {T_{UU}^{scat}} $ is the mean flux given by
\ba
\bk {T_{UU}^{scat}} &=& \expect{0, in} {\left( \partial_U \phi^\dagger
\partial_U \phi +
\partial_U \phi \partial_U \phi^\dagger \right) }{0, in} -
\expect{0,out}{\left( \partial_U \phi^\dagger \partial_U \phi +
\partial_U \phi \partial_U \phi^\dagger \right) }{0, out}
\nonumber\\
&=& 2 \int_0^{\infty} d\la \int_0^{\infty} d\la^\p {\sqrt{\la \la^\p} \over
2\pi} \nonumber\\
&& \quad\quad\left[ e^{-i( \la -\la^\p)U }\left( \int_0^\infty d\om
\bt^*_{\om, \la}
 \bt_{\om, \la^\p}
\right)
- \mbox{Re} \left( e^{-i (\la + \la^\p)U}
\int_0^\infty d\om \al_{\om, \la} \bt^*_{\om, \la^\p} \right) \right]\quad
\label{Tuuq}
\ea
where the Minkowski zero point energy has been subtracted.
Upon comparaison with the classical flux carried by the scattered mode
$\varphi_{\om}^{scat}$,
eq. (\ref{Tuu}),
one sees that in quantum mechanics, the mean flux, i.e. the quantum averaged
flux, is given by the
sum of the classical excesses of the each incoming mode of amplitude squared
$\vert A_\om \vert^2 = \hbar =1 $.
 Indeed the first term of eq. (\ref{Tuu}) when summed over all positive $\om$
gives the zero point energy plus the first term of eq. (\ref{Tuuq}).

Furthermore, as pointed out in ref. \cite{DF},
this mean flux can be expressed in terms of the local properties of the
trajectory $V= V_{cl}(U)$ only. Using Green (Wightman) functions given by
\ba
\expect{0,in} { \phi^\dagger(x) \phi(x')}{0,in}&=&\int_0^\infty {d\om \over
4\pi \om}
\left[ e^{i \om (V'-V)} +e^{i\om \left[ V_{cl}(U')-V_{cl}(U)\right]} \right]
\nonumber\\
\expect{0,out} { \phi^\dagger(x) \phi(x')}{0,out}&=&\int_0^\infty {d\la \over
4\pi \la}
\left[ e^{i \la \left[ U_{cl}(V')-U_{cl}(V)\right]} + e^{i\la (U'-U)} \right]
\label{wight}
\ea
one obtains
\ba
\bk {T_{UU}^{scat}} &=& 2\ \mbox{lim}_{U^\p \to U}
\left[ \expect{0, in} { \partial_{U^\p} \phi^\dagger \partial_U \phi}{0, in} -
\expect{0, out} { \partial_{U^\p} \phi^\dagger \partial_U \phi}{0, out} \right]
\nonumber\\
&=& 2\ \mbox{lim}_{U^\p \to U}
 -{1 \over 4\pi}\partial_{U^\p} \partial_U \left [ \mbox{ln} \vert V_{cl}(U^\p)
- V_{cl}(U) \vert -
\mbox{ln} \vert U^\p - U \vert \right]
\nonumber\\
&=& -{1 \over 6 \pi} \left[  \left( {dV_{cl} \over dU }\right)^{1/2}
\partial_U^2
\left( {dV_{cl} \over dU }\right)^{-1/2}  \right]
\label{localfl}
\ea
Very important is the fact that this equation can be rewritten as
\be
\bk {T_{UU}^{scat}} = {1 \over 12 \pi} \left[ \left(
{ d^2V_{cl}\over dU^2} \right) \left( {dV_{cl} \over dU }\right)^{-1} \right]^2
-{1 \over 6 \pi} \partial_U \left[ \left(
{d^2V_{cl}\over dU^2 }\right) \left( {dV_{cl} \over dU} \right)^{-1} \right]
\label{separ}
\ee
Only the first term, positive definite, contributes to $\bk{H}$ of eq.
(\ref{meanen}).
Indeed for asymptotically inertial trajectories, the second term integrates
to zero. Similarly, the second term of the decomposition of the r.h.s. in
eq. (\ref{Tuuq}) intregrates also to zero.
This latter comes from interferences betwen final states with differents
particle numbers. We recall that $T_{UU}$ contains fluctuating terms
made of products of $ab$ or $a^\dagger b^\dagger$. These products
of operators give  rise to the second term of eq. (\ref{Tuuq}).
One sees therefore that the {\it mean} flux encodes contributions from
interferences which find their origin in the rewriting of the in-vacuum in
terms
of out-states.  Notice the similarity between
the decomposition of eq. (\ref{Tuuq}) and the one of eq. (\ref{separ}). In both
cases,
the first term is positive definite and the second one gives no contribution to
the total energy.

We now mention
some properties of the fluctuations of the out-particle content of
the in-vacuum. There are many ways to describe these
fluctuations.
For instance, one can pick a particular
final state $\ket{\psi}$ and inquire into the probability to find it
on ${\cal I^+}$. This probability is given by
\be
P_{\psi}= \langle 0, in \ket{\psi}\bra{\psi} 0,in \rangle
\label{P}
\ee
Notice that the specification of the final state needs not to be complete.
Indeed one can partially fix the final state by using projectors
on subspaces of the Fock space.
In order to display the correlations among constituents of each pair of quanta,
we now
consider these partial specifications. The reason for this kinematical
exercise is to compare the correlations in this model with the ones
that we shall find in the dynamical model. We shall see that the
notion of "partner" differs significantly since, in the dynamical version,
the field configurations are correlated to the mirror's state as well.

We choose to specify only the particle content of the final state
and we trace over the states of
anti-particle sector. The selected particle is described by the state
\be
\ket{\bar \la} =  \int_0^\infty \! d\la\
f_\la a_\la^{\dagger} \ket{0_a, out}
\label{sstate}
\ee
where $\ket{0_a, out}$ is the particle sector of the out vacuum
on which the $a$ operators act. The projector which specifies that there is
this particle
and which does not specify the state of the antiparticle sector is\cite{MaPa}
\be
\Pi_{\bar \la} = {\mbox{ I}}_{b} \otimes \int_0^\infty \! d\la\
f_\la a_\la^{\dagger} \ket{0_a, out} \bra{0_a, out}
\int_0^\infty \! d\la^\p\
f_{\la^\p}^* a_{\la^\p}
\label{Th}
\end{equation}
where ${\mbox{ I}}_{b}$ is the operator unity acting on the
anti-particle $b$-sector of the final state.
A simple algebra gives the probability to find this state
\be
P_{\bar \la}=\expect{ 0, in}{\Pi_{\bar \la}}{0, in} = {1 \over Z}
\int_0^\infty d\la \vert f_\la \vert ^2 \ \vert \int_0^\infty
 d\om \bt_{\om, \la} \al_{\om , \la}^{-1} \vert ^2
\label{Pi}
\ee
where $\al_{\om , \la}^{-1} $ is the inverse matrix of $\al_{\om \la}$.
The state of the anti-particle correlated to
the state $\ket{\bar \la}$ is defined by $ \ket{\bar \la} \ket{partner,\ {\bar
\la}} =
\Pi_{\bar \la} \ket{0, in}$ and given by
\be
\ket{partner,\ {\bar \la}} = \int_0^\infty d\la  f_\la^* \int_0^\infty
d\om \bt_{\om, \la}^* (\al_{\om , \la}^{-1})^*\ b_\la^\dagger \ket{0_b, out}
\label{partner}
\ee
Owing to the correlations among field configurations in the in-vacuum,
the "partner" is therefore uniquely determined by the $f_\la$'s
but its wave function is not the complex conjugate of the particle one.

If one selects a two particle state
made of orthogonal wave packets and if one still traces over the antiparticle
sector,
 the probability to find this state is the product of
the probabilities to find each particle separately. One has thus production of
statistically independent
pairs of Minkowski quanta whose constituents are perfectly correlated in phase
and amplitude.

Since many final outcomes are possible, the flux of energy fluctuates as well
since it
fluctuates accordingly.
To describe these correlated fluctuations, it is appropriate to evaluate the
following {\it off diagonal} matrix elements of $T_{UU}$\cite{MaPa}\cite{GO}
 \be
\bk{ T_{UU}}_{\Pi} ={ \expect{0, in}{\Pi \ T_{UU}}{0,in} \over
\expect{0, in}{\Pi}{0,in} }
\label{w}
\ee
associated to each final outcome specified by the projector $\Pi$.
This matrix element can be obtained by decomposing the mean flux as
\ba
\langle 0, in \vert T_{UU} \vert 0, in \rangle &=&
\langle 0, in \vert \sum_i \Pi_i T_{UU} \vert 0, in \rangle
\nonumber\\
&=& \sum_i P_i {
\langle 0, in  \vert  \Pi_i T_{UU} \vert 0, in \rangle
\over
\langle 0, in \vert \Pi_i \vert 0, in \rangle  }
\label{r1}
\ea
where the projectors $\Pi_i$ form a complete set which decomposes the
unit operator I as $\sum_i \Pi_i  =$I.
Eq. (\ref{r1}) decomposes the mean flux according to the final state as the
sum of the products of the probability
of finding a particular out come times the matrix element
associated with this outcome. Each matrix element has therefore the
interpretation
of the conditional value of the flux knowing that the initial state is $\ket{0,
in}$ and
the final one $\Pi_i \ket {0, in}$;
see refs. \cite{MaPa}\cite{BMPPS}\cite{GO} for more details. Let us mention
only that
owing to the perfect reflection on $V_{cl}(U)$, one can also calculate the
conditional value of $T_{VV}$ before scattering by the mirror: $
\bk{ T_{VV}}_{\Pi}=\bk{ T_{UU}}_{\Pi} (dV_{cl}/dU)^{-2}$.
Thus $\bk{ T_{\mu \nu}}_{\Pi}$ extends from
$\cal I^+$ back to ${\cal I^-}$ no matter what are the frequencies involved.
Notice that these properties also hold in the black hole case precisely within
the context
 of Hawking's hypothesis of free propagation in a fixed
 background\cite{EMP}\cite{Verl3}. In the next sections, we shall consider
similar off diagonal matrix elements of $T_{\mu \nu}$ in order to isolate the
modifications of these correlations
induced by the recoils of the dynamical mirror.
We shall prove that the correlations to the past are inevitably
washed out  once the characteristic energy
of the fluctuations involved in the matrix element approaches  a certain
function of
the rest mass of the mirror thereby restricting the validity of the results
derived using the
background field approximation.

\section{The Uniformly Accelerating Mirror}

In order to illustrate the relations between
the fluctuating properties of the radiation, the particle concept and the mean
flux,
we analyze in detail the flux emitted by a mirror in constant acceleration.
This case is simple enough to obtain analytical results
but nevertheless possesses intriguing finely tuned interferences which, for
instance,
lead to the vanishing of the mean flux.
We shall generalize the scattering to partially transmitting mirrors
and show that the radiation emitted by these mirrors is closely related to the
radiation emitted by an accelerated oscillator heated by Unruh
effect\cite{Unr}.
In Section 5, we shall reexamine the accelerated mirror with our dynamical
model
and determine the nature of the modifications induced by
 the mirror's dynamics. These put severe restrictions on the validity of the DF
model.

Constant acceleration $a$ means that the trajectory of the mirror satisfies
$V_{cl} (U)= -1/a^2U$. We
put the mirror on the left quadrant: $V<0$, $U>0$. The trajectory
of the mirror does not star at $i^-$ nor end at $i^+$. Indeed, for $t \to
-\infty$,  $U \to 0^+$
and for $t \to \infty$,  $ V \to 0^-$. This fact  justifies or even implies, as
we
shall see, consideration of scattering from both sides from an asymptotically
"transparent mirror".

To explore the properties of the scattering, we first consider the mean flux
emitted to the
right, see eq. (\ref{localfl}). Since $ (dV_{cl}/dU)^{-1/2} =aU$,
$\bk{T_{UU}^{scat}}$
 vanishes\cite{BD}\cite{Grov}.
Nevertheless the field configurations are scattered, hence some Minkowski
quanta are
produced. This can be seen from the non-vanishing character of the $\bt$
coefficients
of the Bogoliubov transformation
\ba
\al_{\om, \la}&=& - \int_0^\infty {dU \over 2\pi} {e^{i\la U} e^{i\om/a^2U}
\over \sqrt{\om/\la} }
= { - 1\over a \pi} K_1\left[ 2i ({\om \la \over a^2})^{1/2} \right]
\nonumber\\
\bt_{\om, \la}&=& - \int_0^\infty {dU \over 2\pi} { e^{-i\la U} e^{i\om/a^2U}
\over \sqrt{\om/\la} }
={ i \over a \pi}  K_1\left[ 2 ({\om \la \over a^2})^{1/2} \right]
\label{K}
\ea
where $K_ 1\left[ z \right]$ is a modified Bessel function, see ref. \cite{BD}.
One then obtains that the total energy, eq. (\ref{meanen}),
associated with those creations diverges.
To have simultaneously a vanishing local flux and a diverging total energy
whose origin comes from singular transcients is a common feature of the
radiation
emitted by accelerated systems when the recoils are completely neglected:
 see \cite{Boul} for the radiation emitted by a classical charge
and \cite{grove}\cite{Unruh92}\cite{AM}\cite{MaPa} for the emission associated
with the
Unruh effect. In each case, one can decompose the mean flux in two terms as in
eq. (\ref{separ}). One finds that the first term describes
a positive constant flux in the local frame (in our case one has:
$\bk{T_{uu}}^{first\ term}=a^2/6 \pi$ where $au=\ln aU$ is the local coordinate
at rest, see below eq. (\ref{rcoord}))
and that the second (negative) term, properly regulated, integrates to zero.
Hence the transcients have a global character. Indeed they encode the total
number of particle created
which is a linear function of the lapse of proper time during which the
acceleration is constant\cite{MaPa}.

Nevertheless, it is not clear how to handle correctly
the fact that the in-modes and the scattered modes are discontinuous
on $U=0$. Therefore it is appropriate to gain some insight by computing
the two point Green function on ${\cal I^+}$. Since the mirror reaches
$V=0$ at $t=\infty$, on the right of it, ${\cal I^+}$ is now the union of $U=
\infty,\ V> 0$ and
$V=\infty$ all $U$'s. Near ${\cal I^+}$, the Green function contains three
types of terms
\ba
{\expect{0, in}{\phi^\dagger(x') \phi(x)}{0, in} \over
- 4\pi} &=& \theta(-U')\theta(-U) \ln \vert U' - U\vert
+\theta(V') \theta(V) \ln \vert V' - V \vert
\nonumber\\
&&+\theta(U')\theta(U) \ln \vert V_{cl}(U') - V_{cl}(U) \vert
\nonumber\\
&&+\theta(V')\theta(U) \ln \vert V- V_{cl}(U) \vert
+\theta(U')\theta(V) \ln \vert V_{cl}(U') - V \vert
\ \ \quad
\label{GF}
\ea
The first two terms give the unscattered propagator in the regions $U<0$ all
$V$'s and
$V>0$ late $U$'s. The second line describes the $U$-part of the scattered field
configurations
as in eq. (\ref{wight}). The third term encodes the correlations set up on
$U=-\infty$ between
positive and negative $V$'s which have been split away from each other owing to
the
asymptotic behavior of the trajectory at late times.
In addition to these unusual correlations,  eq. (\ref{GF})
exhibits also clearly the {\it absence } of correlations\footnote{This absence
of correlations is directly
related to the presence of the
above mentioned singular transcients which encode an infinite energy. This
double aspect was recently
found in the context of black hole evaporation in 2-D dilatonic gravity, see
ref. \cite{Strom}. As suggested
in this reference, the origin of this singular behavior can be imputed to the
use of the
background field approximation. This
will be explicitly proven in Section 5 in the case of the accelerating mirror.
} between positive and
negative $U$'s.

In order to understand the particle content of both aspects, it is very useful
to work with Rindler modes, eigen modes of $iV\partial_V=\nu$. This is because
$V_{cl}= -1/a^2 U$ becomes static in Rindler coordinates given by
\ba
av_L&=& \theta(-V) \ln (-aV )\quad\quad av_R=\theta(V) \ln (aV) \nonumber\\
au_L&=& \theta(U) \ln (aU) \quad\quad au_R= \theta(-U) \ln (-aU)
\label{rcoord}
\ea
Indeed the trajectory reads $u_L=v_L$. Therefore the scattering conserves the
Rindler energy $\nu=
i\partial_v$
and the Bogoliubov transformation relating Rindler modes is diagonal in $\nu$.

The $V$-part of the in-vacuum is vacuum with respect to the
"Unruh"-modes\cite{Unr}
\ba
\varphi_{\nu, V} (V) &=& \al_\nu \varphi_{\nu, R, V} + \bt_\nu \varphi^*_{\nu,
L, V}
\nonumber\\
\varphi_{-\nu, V} (V) &=& \al_\nu \varphi_{\nu, L, V} + \bt_\nu \varphi^*_{\nu,
R, V}
\label{unr}
\ea
where $\nu >0$, where the Bogoliubov coefficients satisfy $\vert \bt_\nu
/\al_\nu \vert ^2 = e^{-2 \pi \nu/a}$
 and where the Rindler modes are\cite{Full}\cite{BD}
\be
\varphi_{\nu, L, V} = \theta(-V) {e^{-i\nu v_L} \over 4\pi \nu}\ ,\quad
\varphi_{\nu, R, V} = \theta(V) {e^{-i\nu v_R} \over 4\pi \nu}
\label{rmodev}
\ee
The $U$-part of the in-vacuum, for $U<0$, is a thermal state with temperature
$a/2\pi$ of Rindler modes
defined by
\be
\varphi_{\nu, R, U} = \theta(-U) {e^{-i\nu u_R} \over 4\pi \nu}
\label{rindl}
\ee
The in vacuum can be expressed in terms of the Rindler operators $a_{\nu, R,
V},\ a_{\nu, L, V}$ and
$a_{\nu, R, U} $ associated with the Rindler modes together with the
corresponding $b$ operators acting on the
Rindler vacuum $\ket{0,R}_V \ket{0,L}_V \ket{0,R}_U \ket{0,L}_U $ as
\ba
\ket{0,in} &=&
\prod_\nu e^{arctanh(e^{-\pi \nu})\left[(a^\dagger_{\nu, R, V}b^\dagger_{\nu,
L, V}+
b^\dagger_{\nu, R, V}a^\dagger_{\nu, L, V} ) +(h.c.)\right]} \ket{0,R}_V
\ket{0,L}_V \nonumber\\
&&\otimes {\mbox{Tr}}_{(a_{\nu, L, U},\ b_{\nu, L, U})}
e^{arctanh(e^{-\pi \nu})\left[(a^\dagger_{\nu, L, U}b^\dagger_{\nu, L, U}+
b^\dagger_{\nu, R, U}a^\dagger_{\nu, L, U} ) +(h.c.)\right]} \ket{0,R}_U
\ket{0,L}_U \ \ \
\label{inv}
\ea
where we have introduced the operators $a_{\nu, L, U}$ and $ b_{\nu, L, U}$ to
represent the thermal trace
for the right $U$ sector.

The $U$ part of the out -vacuum is vacuum with respect to the out-modes given
by
the $U$ version of the Unruh modes displayed in eq. (\ref{unr}), i.e. with $U$
replacing $V$.
The $V$ part of the out-vacuum defined for $V>0$ on $U=\infty$ can be
represented as
was represented the $U$-part of the in vacuum, i.e. by a thermal density matrix
of Right Rindler modes given
in eq. (\ref{rmodev}). Similarly, the out-vacuum can be expressed in terms of
Rindler states
as in eq. (\ref{inv}), with the substitution of $V$ by $U$.

The scattering matrix in Rindler modes is trivial: it replaces $\varphi_{\nu,
L, V} $ by
$-\varphi_{\nu, L, U} $. Therefore, the
scattered Unruh modes are
\ba
\varphi^{scat}_{\nu, V} (U, V) &=&
\al_\nu \varphi_{\nu, R, V} - \bt_\nu \varphi^*_{\nu, L, U}
\nonumber\\&=& \al_\nu \varphi_{\nu, R, V} - \bt_\nu ( \al_\nu \varphi^*_{-\nu,
U} -
\bt_\nu \varphi_{\nu, U})
\nonumber\\
\varphi^{scat}_{-\nu, V} (U,V) &=& -\al_\nu \varphi_{\nu, L, U}+ \bt_\nu
\varphi^*_{\nu, R, V}
\nonumber\\
&=& -\al_\nu (\al_\nu \varphi_{-\nu, U}- \bt_\nu \varphi^*_{\nu, U}) + \bt_\nu
\varphi^*_{\nu, R, V}
\label{scatr}
\ea
In the second equalities we have expressed the Rindler $U$ modes in the r.h.s.
in terms of
Unruh modes.

{}From eq. (\ref{scatr}), one sees that the "partner", see eq. (\ref{partner}),
of a particle
described by a wave packet of $\varphi_{-\nu, U}$ has
two branches. One reaches $V=\infty$ and is described by a superposition of
$\varphi^*_{\nu, U}$ modes
and the other one is a $V$ mode which reaches $U=\infty$. Very important is the
fact that it
 is described by a Rindler mode $\varphi^*_{\nu, R, V}$ since it exists only
$V$.
Therefore this part of the partner wave function
is not a Minkowski quantum and we have not pair production of Minkowski quanta.
 See refs. \cite{Wil}\cite{Carl} for a discussion of the correlations between
asymptotic quanta and "vacuum configurations".
I put quotation marks to emphasize the unusual aspects of these vacuum
configurations
since they are not expressible in terms of Minkowski modes globally defined.
In order to palliate this cumbersome situation it is appropriate to decouple
adiabatically, i.e. with respect to a $1/a$ proper time lapse,
 the mirror
from the field configurations for asymptotic times $t\to \pm \infty$.
Then, the asymptotic field configurations can be safely and
correctly decomposed in terms of globally defined Minkowski modes
both in the past, on $V=-\infty$ all $U$'s,  and  the future, on $U=\infty$ all
$V$'s.
It is very simple to perform this adiabatic decoupling in our case: it suffices
to rewrite the $V$ part of the scattered modes in terms of Unruh modes, eq.
(\ref{unr})
 which are globally defined.
The adiabaticity is precisely what legitimates this simple
procedure\cite{MaPa}.
Thus the operators $a_{\nu, L, V}$ which were introduced only to represent,
through a trace,
the $V$ part
of the out vacuum as a thermal distribution of $R$ quanta, are now treated on
the same
footing as the operators $a_{\nu, R, V}$ and truly represent the field
configurations on the other
side of the mirror. In terms of the Unruh $V$ modes, the relation between the
scattered modes
and the out-modes globally defined are, see eqs. (\ref{scatr}),
\ba
\varphi^{scat}_{\nu, V}&=&(1 + \bt_\nu^2 T^*_\nu)\varphi_{\nu, V}-
\al_\nu \bt_\nu T^*_\nu \varphi^*_{-\nu, V} + \al_\nu \bt_\nu T^*_\nu
\varphi_{-\nu, U}^*
-\bt_\nu^2 T^*_\nu  \varphi_{\nu, U}
\nonumber\\ \varphi^{scat}_{-\nu, V}&=&
(1 -\al_\nu^2 T_\nu )\varphi_{-\nu, V} + \al_\nu \bt_\nu T_\nu \varphi^*_{\nu,
V} -
\al_\nu \bt_\nu T_\nu \varphi_{\nu, U}^* +
\al_\nu^2 T_\nu \varphi_{-\nu, U}
\label{scatt}
\ea
 together with the equations for the scattered $U$-modes
given by the same equations with $U$ and $V$
interchanged. Indeed if one decouples
the field configurations at $t=\infty$, the $U$ modes are also scattered.
We have introduced the complex coefficients $T_\nu$ to generalize to the case
of partially
 reflecting mirrors. The coefficients $T_\nu$ satisfy the unitary relation:
Re$T_\nu =
\vert T_\nu \vert ^2$.
Total reflection is given by $T_\nu =1$. Of course for $T_\nu=0$, there is no
 production since there is no scattering.
One has now a "normal" pair creation phenomenon in which the asymptotic quanta
are
all free Minkowski quanta. The pairs are made out of $U$-$U$, $U$-$V$ and
$V$-$V$ quanta.
Therefore one sees
that the adiabatic decoupling has replaced the correlations between Minkowski
quanta
and vacuum field
configurations\cite{Wil} by pair production occurring on both sides of the
mirror.
It should be noted that the decoupling has been introduced here in a rather
{\it ad hoc} way.
Nevertheless, this procedure will
be justified in the next sections since we shall see that it will
come up automatically, in perturbation theory, when we shall compute the $S$
matrix elements.
Note that one can also refuse to switch on and off asymptotically
the coupling to the mirror. In that case,
one has to analyze the scattering in terms of asymptotic states which are
not the free Minkowski modes but which are instead characterized by a "final
state interaction" with the
mirror.

The adiabatic decoupling is already justified by  following remark.
The partial scattering from the mirror, eq. (\ref{scatt}), is very similar to
the
scattering of light like quanta induced by Unruh effect\cite{Unr},
 i.e.  the fact that the inner degrees of freedom of an
accelerated system thermalize with temperature $a/2 \pi$. Indeed one can
 pass continuously from one case to the other. Take for instance the
model of the oscillator introduced by Raine, Sciama and Grove\cite{RSG}, see
also \cite{MPB}.
In that model the oscillator is coupled to the derivative of the
field by the hamiltonian $
H_{int}=e \int d\tau i [\dot \phi(\tau)- \dot \phi^\dagger(\tau) ]
(A e^{-i\mu\tau} + A^\dagger e^{i\mu\tau})$. $A$ ($A^\dagger$) is the lowering
(raising) operator
 and $\mu$ is the energy gap between two states.
Unruh's analysis\cite{Unr} shows that this accelerated system
thermalizes in Minkowski vacuum with temperature $a/2\pi$, i.e.
 that the equilibrium probabilities to find the oscillator in its
nth excited state satisfy $P_{n}=e^{-2 \pi n \mu /a}P_{gr.}$, where $P_{gr.}$
is the
probability to find the oscillator in its ground state. Upon eliminating the
oscillator
operators, one can express the
 scattered Rindler modes in terms of the out Rindler modes. One
obtains\cite{MaPa}
\be
\varphi^{scat} _{ \nu,L,U} =  \varphi _{ \nu,L,U} (1+ i{e\over 2} \psi _{
\nu})+
(i{e\over 2} \psi _{ \nu})  \varphi _{
\nu,L,V}
\label{neuf}
\ee
where $\psi_\nu$ is the oscillator's propagator given by
\be
\psi _{ \nu} = { e\nu \over  \mu^2 -  \nu ^2 -i e^2  \nu /2}
\label{sept}
\ee
Therefore $T_{\nu}= i e \psi _{ \nu} /2 $.
 Furthermore upon taking the limit $e \to \infty$, one obtains total
reflection.

By virtue of this mapping, all properties of $T_{\mu \nu}$ including
the conditional value of $T_{\mu \nu}$, eq. (\ref{w}), are
in strict analogy with the properties obtained from Unruh effect. We therefore
refer to
\cite{MaPa} for a detailed analysis of the the fluctuating properties of
$T_{\mu \nu}$.
Here, we shall only present
an interesting link between the fluxes emitted by an accelerated oscillator
and by the accelerated mirror which sheds light on the nature of the
mirror when viewed as a quantum system.

In the low coupling limit, i.e for $e^2<<\mu$, the norm of the
oscillator's propagator $\psi_{\nu}$ tends to $\delta(\mu -\nu)+ \delta(\mu +
\nu)$.
Then, once thermal equilibrium is achieved,
the mean flux emitted by the oscillator is, see
\cite{UnrW}\cite{grove}\cite{RSG}\cite{MaPa},
\ba
\bk{T_{UU}}^{atom}_{\mu}
&=& \theta(U) \ 2 e^2 \mu \int_0^{\infty} d\la \int_0^{\infty} d\la^\p
{\sqrt{\la \la^\p} \over 2\pi}
 \nonumber\\
&& \quad {1 \over 1- e^{-2 \pi \mu /a}} \left[ e^{-i( \la -\la^\p)U } \bt^{L*}
_{\mu, \la}
 \bt^{L}_{\mu, \la^\p}
- \mbox{Re} \left( e^{-i (\la + \la^\p)U} \al^L_{\mu, \la} \bt^{L*}_{\mu,
\la^\p} \right)
\right]\quad
 \nonumber\\
&& \quad +{1 \over e^{2 \pi \mu /a}- 1} \left[ e^{-i( \la -\la^\p)U }  \al^L
_{\mu, \la}
 \al^{L*}_{\mu, \la^\p}
- \mbox{Re} \left( e^{-i (\la + \la^\p)U} \bt^{L*}_{\mu, \la} \al^{L}_{\mu,
\la^\p} \right)
\right]\quad
\label{Tuuqatom}
\ea
where $ \bt^{L}_{\mu, \la^\p}$ and $ \al^{L}_{\mu, \la}$ are the Bogoliubov
coefficients between
Rindler modes of frequency $\mu$ and the Minkowski modes of frequency $\la$.
The first line comes from the excitation process and the second line from the
deexcitation.
The thermal factors multiplying the brackets are respectively the probabilities
$P_{gr.}$ and $P_{1}$. The overall
factor $\mu$ comes from the derivative coupling
and the relativistic normalization of the modes.

Thus upon summing $\bk{T_{UU}}^{atom}_{\mu}/\mu$  over all $\mu$'s,
one obtains $e^2$ times the flux reflected by the mirror
reexpressed in terms of Unruh modes, eq. (\ref{unr}),
instead of the usual Minkowski modes as done in eq.
(\ref{Tuuq}). Indeed in that equation,
the sum over all positive $\om$ can be replaced by any other complete set of
modes of
positive in frequency. Therefore,
the mirror acts as a collection of weakly coupled
(to the time derivative of the field) oscillators
 whose density of probability of finding the
frequency $\mu$ is given by $d\mu/\mu$.

It is now obvious that the (mean) total number of Minkowski quanta emitted by
the mirror
increases with the lapse of proper time during which the mirror interacts with
the field\cite{MaPa}, since
each excitation or deexcitation process leads to the production of a Minkowski
quantum in spite of the
fact that the {\it local} flux, $\bk{T_{UU}}^{atom}_{\mu}$, vanishes for all
$\mu$ in the
intermediate region. In Section 5, we shall see that the vanishing character of
$\bk{T_{UU}}$ as well as
its singular behavior at $U=0$ are both directly attributable to the classical
and unaffected character
of the mirror's trajectory. Indeed, upon taking into account the recoils of the
mirror, we shall see that
the local repartition of the energy content of the emitted quanta is
completely modified after a certain lapse of proper time
even though neither the
mean number of quanta emitted nor their energy is changed.

\section{The Dynamical mirror}

We shall introduce and analyze a dynamical version of the DF model in the
simplest case, that
of the inertial mirror. A dynamical non inertial mirror will be considered in
the next section.
We shall proceed in two steps. First we shall introduce a self interacting
model without
dynamics. The reasons for this intermediate step are the following:

\noindent
1) The intermediate model gives back the DF model upon resumming the Born
series.

\noindent
2) The perturbative matrix elements of the intermediate model coincide with
  the large mass limit of the matrix elements of dynamical model we shall
discuss afterwards.

Therefore, the dynamical model is related to the DF model by a well defined
double procedure:
 by a limit of large mass and by resummation of perturbative effects. In
addition to the
large mass limit, we shall also describe the other limit where the
energy of the incident quantum approaches the mirror's mass. In this case, we
shall see that
the mass of the mirror acts as a cut off for the reflected energy.

{}From the equality of the matrix elements in the large mass limit, (point 2),
it is tempting to conclude
that one should obtain, in this limit, the same
expression as the one obtained in the DF
or the intermediate model whatever is the quantity calculated. This is not the
case.
The reason is
that the dynamical model contains
a third dimensionful quantity, in addition to the frequency of the scattered
light
and the mass of the
mirror, namely the spread of the wave function which characterizes the position
of the mirror.
Together with the momentum transfer to the mirror upon
reflection, this spread intervenes in decoherence effects when the frequency
of the scattered light is comparable or bigger than its inverse
but nevertheless much smaller than the rest mass of the mirror. To describe
this
decoherence we shall study the mean scattered flux when the incident radiation
state
is described by a pure state which contains states with different particle
number. In that case,
the mean flux can be decomposed into two terms as in eqs. (\ref{Tuu},
\ref{Tuuq}). We shall see
that
each term is differently affected by the recoils and that these modifications
lead to decoherence effects.

We start by analyzing the intermediate model in order to prove point 1).
The equation which replace eqs. (\ref{one}, \ref{two}) in the inertial case
 and which defines the intermediate model
is
\be
\left[ \partial_t^2 -\partial_z^2 \right] \phi(t,z) = g \delta(z) \
2i\partial_t \phi(t,z)
\label{model2}
\ee
where $g$ is a dimensionless coupling constant and $\delta(z)$ is a $\delta$ of
Dirac which specifies that the scatterer sits for all times on $z=0$.

Our goal is to prove that in the large (compare to 1) $g$ limit, the scattering
of light
induced by the r.h.s of eq. (\ref{model2}) gives back the DF model, i.e. gives
back the coefficients
$\gamma_{\om, \la}$ defined in eqs. (\ref{seven}, \ref{eight}).
This is an easy task since eq. (\ref{model2}) is linear in $\phi$.
Therefore, in classical as well as in quantum mechanics, the solution can be
expressed as
\be
\phi(t,z) = \phi^{in}(t, z) +\int_{-\infty}^{\infty} dt'
\int_{-\infty}^{\infty} dz'
G^{ret}(t', z'; t, z) g \delta(z') \ 2i\partial_t \phi(t', z')
\label{eqlin}
\ee
where $\phi^{in}$ satisfies the free, $g=0$, d'Alembertian and
where $G^{ret}$ is the retarded Green function. In Fourier decomposition is it
given by
\be
G^{ret}(t', z'; t, z) =
\int_{-\infty}^{\infty} d\om \int_{-\infty}^{\infty} dk
{1 \over 4 \pi^2} {e^{i\om(t' -t ) -i k (z' -z)} \over (\om + i \e)^2 - k^2}
\quad (=0\ \mbox{for} \ t'>t)
\label{ret}
\ee

Since the location of the mirror is static it is appropriate to work at fixed
energy\cite{mt}.
One defines
\be
\phi_\om(z) = \int_{-\infty}^{\infty} {dt \over 2\pi} \phi(t,z) e^{i\om t}
\label{phiom}
\ee
In terms of these Fourier components, eq. (\ref{eqlin}) becomes
\ba
\phi_\om(z) &=& \phi_\om^{in} (z) + ig \phi_\om(z=0) (2i\om) \int {dk \over
2\pi}
{e^{ikz} \over (\om + i \e)^2 - k^2}
\nonumber\\
 &=& \phi_\om^{in} (z) - ig \phi_\om(0) {e^{-i\om \vert z \vert} }
\label{eqres}
\ea
Similarly one can express the Heisenberg (interacting) field $\phi$ in terms of
the
out-field $\phi^{out}$ and the {\it advanced} Green function given by eq.
(\ref{ret}) with
$\e =0^-$. One gets
\be
\phi_\om(z)=
\phi_\om^{out} (z) + ig \phi_\om(0) {e^{i\om \vert z \vert} }
\label{outm}
\ee

Therefore one can eliminate the interacting fields $\phi_\om(z)$ and
$\phi_\om(z=0)$ and one obtains
\be
\phi_\om^{out} (z)= \phi_\om^{in} (z)
+ ig  \left( {1 \over 1 -i g } \right) \phi_\om^{in} (z=0) \ 2\cos (\om z)
\label{sol}
\ee
By expressing the operator $\phi_\om^{in} (z)$ is terms of the usual
Minkowski operators $a_{\om, i}$ where $i = U$ or $V$ and $\om= \vert k \vert$
is the energy,
one gets
\be
a_{\om, U}^{out} = a_{\om , U}^{in} \left( 1 + {ig \over 1 - ig} \right) +
a_{\om , V}^{in}
\left(  {ig \over 1 - ig} \right)
\label{aa}
\ee
Thus for $g \to \infty$, one obtains pure reflection for an inertial mirror,
i.e.
$a_{\om, U}^{out} = - a_{\om , V}^{in}$ or $\gamma(\om, \la) = -\delta (\om
-\la)$ in eq. (\ref{eight}).
We have therefore proven point 1)  given above. Had we taken $g \delta(z) \phi$
for the r.h.s. of eq.
(\ref{model2}), we would have got, as in ref. \cite{mt}, a frequency dependent
coefficient.

In preparation for point 2), we compute the perturbative matrix elements of
this intermediated model.
In the interacting representation, one has
\be
S_{\om, \la}=\bra{0} a_{\la, j} \ S\  a^\dagger_{\om, i} \ket{0} =
\bra{0} a_{\la, j} \ \mbox{T} e^{-i\int dt H_{int}} \ a^\dagger_{\om, i}
\ket{0}
\label{S1}
\ee
where $\ket{0}$ is Minkowski vacuum,  T
is the time ordered product and $H_{int}$ is the interaction hamiltonian given
by
\be
H_{int}= g \int_{-\infty}^\infty dz \delta(z)
 \phi^\dagger  i\!\lr{\partial_t} \phi = g \int_{-\infty}^\infty dz \delta(z)
J_t(t,z)
\label{inter}
\ee
Using this hamiltonian, the Hamilton equations lead to eq. (\ref{model2}).
To first order in $g$,
the matrix element is
\be
S_{\om, \la} =  \delta (\om -\la) \delta_{ij} - ig
\int_{-\infty}^\infty dt \bra{\la, j} \phi^\dagger i\!\lr {\partial_t} \phi
\ket{\om, i}
 = \delta (\om -\la) \delta_{ij} - ig \ \delta(\om -\la)
\label{mate}
\ee
where we have written momentum conservation as energy conservation times
the $\delta_{ij}$ of Kronecker between $U$ and $V$ sectors.
Notice that the second term, linear in $g$, expresses only energy conservation
as it should do.
Indeed, to first order in $g$, it corresponds to the result of eq. (\ref{aa})
in the interacting picture.

Notice also that one can easily generalize this scattering to non inertial
trajectories.
It suffices to replace the interaction hamiltonian, eq. (\ref{inter}),
by
\be
H_{int} = g \int_{-\infty}^\infty dz
\delta(z_{cl}(t) - z)
\dot x^\mu_{cl}(t)  J_\mu(t, z)
\label{noninert}
\ee
where $\dot  x^\mu_{cl}(t) =dx_{cl}^\mu(t)/dt $ is the tangent vector of the
classical trajectory $z=z_{cl}(t)$ and
$J_\mu$ is the current operator.
In this non inertial case, to first order in $g$, one verifies that the matrix
elements $S_{\om, \la}$
identically give the $\gamma_{\om, \la}$
 coefficients of the DF model, see eq. (\ref{eight}), since these later
are given in terms of matrix elements of the current, see eq. (\ref{four}).
Finally one can also generalize this model to smeared off $\delta$ functions.
One then finds that the matrix elements decrease when the incident frequency
$\om$ is higher than
the frequency content of this smeared function.

We now define the inertial version of our dynamical model for the mirror.
The mirror is described by a charged scalar massif field $\psi$ of mass $M$
coupled to the massless
field $\phi$ by the following interaction hamiltonian
\be
\tilde H_{int} =
\tilde g \int_{-\infty}^{\infty} dz J^\mu_\psi (t, z) J_\mu ^\phi (t, z)
\label{hamilt2}
\ee
This hamiltonian gives a four point interaction through the currents carried by
$\psi$
and $\phi$. (This is why we choose to work with complex fields.)
The Euler-Lagrange equations are
\ba
\left[ \partial_t^2 -\partial_z^2 \right] \phi(t,z) &=&
\tilde g J^\mu_\psi (t, z) \ 2i\partial_\mu \phi(t, z)
\nonumber\\
\left[ \partial_t^2 -\partial_z^2 + M^2 \right] \psi (t, z) &=&
\tilde g J^\mu_\phi(t,z)\ 2i\partial_\mu \psi(t, z)
\label{EL}
\ea

One can verify that in a naive "background field approximation"
this model gives back the intermediate model. Indeed, in the large mass limit
and
upon neglecting recoils, the current carried by the heavy $\psi$ field can be
approximated by a $\dot  x^\mu_{cl}(t) \delta(z_{cl} (t) - z)$
evaluated along its classical trajectory. In this approximation, if $\tilde g
=g$, the hamiltonian
$\tilde H_{int}$ gives back the hamiltonian of the
intermediate model given in eq. (\ref{noninert}).
We shall not pursue in this way but on the contrary we shall work in a pure
quantum mechanical
framework.

Therefore we shall compute the perturbative $S$ matrix elements induced by
$\tilde H_{int}$.
We work in the interacting representation. Then the annihilation and creation
operators
are the usual free Minkowski ones. The operators for the $\phi$ field have
already been given. The
operators for the $\psi$ field are obtained by decomposing the free $\tilde g
=0$ solutions
of eq. (\ref{EL}) in Fourier terms
\be
\psi(t, z) = \int_{-\infty}^{\infty} dp \left[ c_p  {e^{-i(\Om_p t -p z)} \over
\sqrt{4 \pi \Om_p}} +
d^\dagger_p {e^{i(\Om_p t -p z)} \over \sqrt{4 \pi \Om_p}}
\right]
\label{psif}
\ee
where $\Om_p^2 = p^2 + M^2$. The operators $c_p$ annihilate the vacuum state
$\ket{0}_{Mir}$.

The matrix element between an initial state containing an heavy particle of
momentum $p$ and a
lightlike quantum of momentum $k_\om $ and a final state
of momenta $p'$ and $k_\la$ is
\be
\tilde S(p; k_\om ; k_\la) = \bra{\tilde 0} c_{p'} a_{k_\la} \ \tilde S \
a^\dagger_{k_\om}
c^\dagger_p \ket{\tilde 0}
= \bra{\tilde 0} c_{p'} a_{k_\la} \ \mbox{T} e^{-i\int dt \tilde H_{int}}\
a^\dagger_{k_\om}
c^\dagger_p \ket{\tilde 0}
\label{tildeS}
\ee
where $\ket{\tilde 0} = \ket{0}_{Mir} \ket{0}$. To first order in $\tilde g$,
$\tilde S(p; k_\om ; k_\la) $
 is given by (compare with eq. (\ref{mate}))
\ba
\tilde S(p; k_\om ; k_\la) &=&
\delta(p'-p) \delta(\la-\om) \delta_{j i} - i \tilde g
\int_{-\infty}^{\infty} dt \int_{-\infty}^{\infty}  dz
\bra{p'} \psi^\dagger i\!\lr {\partial_\mu} \psi \ket{p} \bra{k_\la}
\phi^\dagger i\!\lr {\partial^\mu}
 \phi \ket{k_\om}
\nonumber\\
&=&
\delta(p'-p) \delta(k_\la-k_\om) - i \tilde g
\delta(p'-p -k_\om + k_\la) \delta(\Om_{p'} + \la - \Om_{p} - \om)\nonumber\\
&&\quad\quad\quad\quad\quad\quad\quad\quad\quad\quad
\left[{( \Om_{p'} +\Om_{p})(\om + \la) -( p'+p)(k_\om+k_\la) \over
4\sqrt{\Om_{p'}\Om_{p}\om \la}}
\right]
\label{deltass}
\ea
The first $\delta$ of the linear term in $\tilde g$
expresses momentum conservation and the second one gives energy conservation.
When the initial momentum $p$ vanishes, this latter gives
\ba
\la&=&\om \quad \mbox{if} \quad i=j
\nonumber\\
\la &=& {\om \over 1 + 2 \om /M}
\quad \mbox{if} \quad i \neq j
\label{scatM}
\ea
As for the passive mirror, there is no energy transfer
 in the case of scattering which conserves $U$-ness or $V$-ness; only a
phase is introduced upon scattering.
On the contrary, in the reflection case,
one recovers energy conservation only for the photons whose energy satisfies
$\om /M << 1$.
In that case the coefficient in bracket in eq. (\ref{deltass}) is unity for all
such small $\om$ and $\la$.
When $\om /M <<1$ but not negligible, the first correction to the energy
relation between $\la$
and $\om$,
eq. (\ref{scatM}), and to the bracket of eq. (\ref{deltass}) is
the non relativistic energy $(2\om)^2 /2M$ of the mirror. Therefore
we have proven point 2) since, in the
limit $M \to \infty$ at fixed $\om,\ \la$ and $p$, $\tilde S(p; k_\om ; k_\la)
= S(k_\om ; k_\la) $,
 the matrix element of eq. (\ref{mate}).

On the contrary, when $\om /M \to \infty$, one finds $\la \to M/2$.
Thus, the finiteness of the mass of the mirror acts as a U.V. cut off for the
reflected frequency
and the discrepancy with the DF model is total. In that respect we mention ref.
\cite{mt2} wherein it is shown how
an U.V. cutoff on the reflection coefficients
eliminates the divergent expressions of the Casimir effect (and therefore the
necessity
of the regularization and the renormalization procedure) since these
frequencies no longer contribute to the Casimir force. Notice also that both
the correction
term $(2\om)^2 /2M$ for small $\om /M$ and the asymptotic behavior $\la \to
M/2$ for large
$\om /M$ are quantum mechanical in nature: reinstating $\hbar$ one has:
$\la = \om - \hbar (2\om)^2 /2M$ and $\la \to \hbar M/2$. In classical
mechanics these terms encode
the nonlinearity of the theory since the scattered frequency is now a function
of the {\it norm}
 of the incident wave.

In view of the fact that in the limit $M \to \infty$ at fixed $\om, \la, p$,
one has $\tilde S(p; k_\om ; k_\la) = S(k_\om ; k_\la) $, it is tempting to
conclude that the results
evaluated in the dynamical model would coincide, in {\it all situations},
 with the ones obtained in this no-recoil model.
This is not the case. There are circumstances in which interferences induced by
momentum
transfer to the mirror upon reflection
ruin the equality of the results. Indeed being related to momentum transfer,
the interferences subsist
in the limit $M\to \infty$ at fixed $\om, \la, p$. Furthermore,
they induce decoherence. To show how and why this happens,
we shall now study the properties of the scattered flux.

\vskip 1. truecm
{\it The scattered flux}

\noindent
Our goal is to show that even in the limit $M \to \infty$, in certain
circumstances,
 expectation values of local operators do not coincide with the corresponding
expectation values evaluated in the no-recoil model because additional
phase factors induced by recoils arise in the expectation values.
To describe the role of these factors we shall compute
the local properties of the reflected flux when the radiation state contains
states with different particle number.  In that case, the mean flux possesses
two terms
as in eqs. (\ref{Tuu}, \ref{Tuuq}). Then we shall see clearly the different
consequencies
of the recoils on each term.
Similar effects will be found in the next section upon computing the properties
of the flux spontaneously produced by a non-inertial mirror.
In both cases, when the momentum transfers are comparable to the inverse spread
of the
wave function of the mirror, the second term of the decomposition of the mean
flux
vanishes.

We work in the limit $\om/M \to 0$ and $ p/M \to 0$. Then the energy of the
radiation
is exactly conserved, see eq. (\ref{scatM}),
 and the mirror has no energy: its wave function does not spread in time but
it still absorbs momentum upon reflecting a quantum. To exhibit how recoils
lead to decoherence, we shall analyze the scattering of the following radiation
state
\be
\ket{\chi}= A \ket{0} + B \int_0^\infty d\om_1 \int_0^\infty d\om_2 \
f_{\om_1} g_{\om_2} a^\dagger_{\om_1, V} a^\dagger_{\om_2, V} \ket{0} =  A
\ket{0} + B \ket{2}
\label{khi}
\ee
where $A$ and $B$ are coefficients such that $\bra{\chi} \chi \rangle =1 $
and where $f_{\om}$ and $g_{\om}$ describe two normalized wave packets which
 have non overlapping frequency range. (This requirement simplifies the
following equations.)
The reason for having chosen $\ket{\chi}$ is that it contains states whose
particle
 number differ by two, hence
there will be a contribution to the mean flux arising from the "interfering"
second term of
 eq. (\ref{Tuuq}). We
shall say that the radiation states decohere when the scattered state of the
coupled system
radiation+mirror is such that this interfering term vanishes owing to the
recoils induced by the
reflection of the quanta building this flux.

Before scattering, the normalized wave function of the mirror is
\be
\ket{Mir} = \int_{-\infty}^\infty dp \  h_p c^\dagger_p \ket{0}_{Mir}
\label{mirror}
\ee
where the momentum profile $h_p$ is $ e^{-p^2/2\sigma^2} /({\pi
\sigma^2})^{1/4}$.
We do not assume that the frequency content of $f_\om$ and $g_\om$ is
negligible
with respect to $\sigma$, in which case one would have a mirror with a well
defined position with
respect to $1/\om$.
On the contrary, we shall see that decoherence occurs when the characteristic
frequencies in $f$ or $g$ are comparable or greather than $\sigma$.

The initial state of the total system is thus $\ket{in} = \ket{Mir}
\ket{\chi}$.
Before studying the scattering, we analyze the properties of the initial flux.
The total energy of the radiation, before scattering, is
\ba
\bk{H_V}&=& \expect{in}{\int_0^\infty d\om \om a^\dagger_{\om, V} a_{\om, V}
}{in} =
\expect{\chi}{\int_0^\infty d\om \om a^\dagger_{\om, V} a_{\om, V} }{\chi}
\nonumber\\
&=&
\vert B \vert ^2 \left[
\int_0^\infty d\om \om \vert f_{\om} \vert ^2 +\int_0^\infty d\om \om \vert
g_{\om} \vert ^2
\right]
\label{Hv2}
\ea
since $\bra{Mir} Mir\rangle =1$. The mean energy is thus the sum of the mean
energies
of each initial quantum times the probability to find it.

The mean initial energy density is not so simple. It is given by
\ba
\bk{T_{VV}} &=& \expect{in}{\partial_V \phi^\dagger \partial_V \phi +\partial_V
\phi
\partial_V \phi^\dagger }{in} =  \expect{\chi}{\partial_V \phi^\dagger
\partial_V \phi +\partial_V \phi
\partial_V \phi^\dagger }{\chi}
\nonumber\\
&=&\vert   B \vert ^2 \expect{2}{\partial_V \phi^\dagger \partial_V  \phi
+\partial_V \phi
\partial_V \phi^\dagger}{2} + 2\mbox{Re} \left[
A^* B \expect{0}{\partial_V \phi^\dagger \partial_V \phi}{2} \right]
\label{Tvvv}
\ea
where $\ket{2}$ designates the second term of eq. (\ref{khi}).
The first term of eq. (\ref{Tvvv}) comes from the diagonal part of the operator
$T_{VV}$. It is given by
\be
\vert   B \vert ^2 \int_0^\infty d\om_1 \int_0^\infty d\om_2
{\sqrt{ \om_1 \om_2} \over 2\pi}e^{-i(\om_1-\om_2)V}
\left[ f_{\om_1}f_{\om_2}^* +g_{\om_1}g_{\om_2}^*
\right]
\label{T1}
\ee
This term is thus the sum of the mean fluxes carried by each wave function
times
the probability to find the particles. Each contribution corresponds exactly to
the first
term of eqs. (\ref{Tuu}, \ref{Tuuq}). (There is no crossing term in
$g^*_{\om_1}f_{\om_2}$ because we have imposed that the frequency ranges
of $f$ and $g$ do not overlap.)

The second term of eq. (\ref{Tvvv})
comes from interferences between states whose particle number differ
by two. It reads
\be
\int_0^\infty d\om_1 \int_0^\infty d\om_2 {- \sqrt{ \om_1 \om_2} \over 2\pi}
\mbox{Re} \left[
e^{-i(\om_1+\om_2)V}  \left(A^* B  f_{\om_1}g_{\om_2} \right)
\right]\ \ \
\label{T2}
\ee
As for the second terms of
 eqs. (\ref{Tuu}, \ref{Tuuq}), when integrated over all $V$'s, it vanishes
 identically. Notice that it depends also of the phase of $A^* B $, i.e. the
relative phase between
the vacuum and the two particle state. We shall now prove that this term is
sensitive to the
above mentioned additional phases induced by the recoils.
To this end, we compute the scattered state and the properties of the scattered
flux. We shall then
compare these properties with the properties before scattering.

To order $\tilde g^2$, the scattered state on $t=\infty$ has the following
structure,
\be
\tilde S \ket{in} = \ket{in} -i \tilde g \ket{Mir' ,1 scat} - \tilde g^2
\ket{Mir'', 2 scat}
\label{struct}
\ee
where $\ket{Mir', 1 scat}$ designates the state of the system on
which the interaction
hamiltonian, eq. (\ref{hamilt2}), has acted only once. In that state only one
particle has been scattered.
The state $\ket{Mir'',
2 scat}$ is obtained by acting twice with this hamiltonian. The relevant part
of this state
is
that part in which both particles have been reflected. Indeed the other parts
will not contribute to the following expressions, at least to order $g^2$.
The explicit expressions of these states are easily obtained in the Golden
Rule approximation since in that case, it is legitimate to neglect the time
ordering appearing in
the $\tilde g^2$ term, see \cite{MaPa}. They will not be explicitized here.
What interests us is the structure of the
scattered flux.

 To order $\tilde g^2$, the reflected mean energy is
\ba
\bk{H_U^{scat}} &=&  \expect{in}{\tilde S^\dagger \
H_U
\ \tilde S}{in} = \tilde g^2  \expect{Mir', 1 scat} {\int_0^\infty d\la \la
a^\dagger_{\la, U}
a_{\la, U} }
{Mir', 1 scat} \nonumber\\
&=& \tilde g^2
\vert B \vert ^2
\left[ \int_0^\infty d\la \la \vert f_{\la} \vert ^2 +\int_0^\infty d\la \la
\vert g_{\la} \vert ^2
\right]
\label{Hv3}
\ea
Note that the twice scattered state $\ket{Mir'',
2 scat}$ does not contribute to $\bk{H_U^{scat}}$. Furthermore, since $H_U$ is
diagonal in $\la$,
only diagonal terms in $p'$  contribute (where $p'$ is
the scattered momentum of the
mirror: $p'=p-2\om$, see eq. (\ref{deltass})).
Thus the scattered mirror wave function factorizes out.
To obtain the third equality, we have used
 the kinematical conditions: $\om_i << M$ and $p << M$ which give
$\la_i=\om_i$.
Thus we see that the reflected energy is simply
the initial energy, eq. (\ref{Hv2}),
times $\tilde g^2$, the probability to be reflected by the mirror. This is true
no matter how
big or small is $\sigma$, the spread of momentum of the initial wave packet of
the mirror, eq. (\ref{mirror}).

The reflected local flux of energy is more complicated and does depend on
$\sigma$.
To order $\tilde g^2$, one finds
\ba
\bk{T^{scat}_{UU}}&=& \tilde g^2 \vert B \vert ^2  \expect{Mir', 1 scat}{
\partial_U \phi^\dagger \partial_U \phi +\partial_U \phi
\partial_U \phi^\dagger }
{Mir', 1 scat} \nonumber\\
&&- \tilde g^2 \ 2\mbox{Re} \left[
A^* B \bra{Mir}\expect{0}{\partial_U \phi^\dagger \partial_U \phi }{Mir'',2
scat} \right]
\label{Tuuscat}
\ea
Straightforward algebra gives both terms. Before giving the resulting
expressions it is
worthwhile to make the following observation. In the second term, upon
evaluating the
matrix element of $T_{UU}$, one finds that the overlap the mirror's unscattered
state with the
twice scattered one gives:  $\int dp h^*_p h_{p-2\om_1 -2 \om_2}=
 e^{-2(\om_1 +\om_2)^2/\sigma^2}$. Therefore if the characteristic frequencies
of $f$ and $g$
are bigger than $\sigma$, this term vanishes. The former interferences between
$\ket{0}$ and $\ket{2}$ have been washed out by the
momentum transfers to the mirror.
 In position terms this washing out may be understood as follows. The
interfering pattern of the incident flux encoded in eq. (\ref{T2}) is
characterized by the
inverse mean frequency $1/\om$. Upon scattering, it is smeared out over the
inverse frequency of the mirror: $1/\sigma$ and therefore it vanishes since
integrated
over {\it all} $U$'s it does so. This simple averaging seems to be directly
attributable to the
broad character of the initial wave of the mirror and would correspond to the
scattering from
a smeared out passive mirror, i.e. by the hamiltonian eq. (\ref{inter}) with
$\delta(z)$ replaced by a
$d(z) = \int dp e^{ipz} h_p$.
This is an overhasty conclusion since for $\om >> 1/\sigma$, the passive mirror
is completely
transparent.

Explicit expressions for both terms of eq. (\ref{Tuuscat}) are listed below.
The first one is
\be
 \tilde g^2 \vert B \vert ^2 \int_0^\infty d\om_1 \int_0^\infty d\om_2
{\sqrt{ \om_1 \om_2} \over 2\pi}e^{-i(\om_1-\om_2)U}
\left( f_{\om_1}f_{\om_2}^*
+g_{\om_1}g_{\om_2}^*\right) \int dp h^*_{p-2\om_2} h_{p-2\om_1} \ \ \ \
\label{firstT}
\ee
As in eq. (\ref{T1}), we have
 used the non-overlapping character of the frequency ranges of $f$ and $g$.
Notice how the overlap of the two scattered states of the mirror enters into
the integrant.
When $\sigma << \om$, its role is to spread out the individual beams of energy
on scales given
by $1/\sigma $.
The second term of eq. (\ref{Tuuscat}) is
\be
-\tilde g^2 \int_0^\infty d\om_1 \int_0^\infty d\om_2 {- \sqrt{ \om_1 \om_2}
\over 2\pi}
\mbox{Re} \left[
e^{-i(\om_1+\om_2)U}
 \left(A^* B  f_{\om_1}g_{\om_2} \int dp h^*_p h_{p-2\om_1 -2 \om_2} \right)
\right]
\label{secondT}
\ee
compare with eq. (\ref{T2}). For obtaining this result we have neglected the
time ordering appearing to this
order in $\tilde g$. It is a legitimate simplification in our case where
$\om/M<< 1$. As
already mentioned the overlap between the two states of the mirror vanishes
when
 $\sigma << \om$.

It is worthwhile to point out that in the limit  $\sigma >> \om$ these
expressions are equal (up to a sign)
to the ones obtained before scattering.
Indeed eq. (\ref{firstT}) coincides
with eq. (\ref{T1}) and eq. (\ref{secondT}) is the opposite of eq. (\ref{T2}).
This sign flip comes from two developments of $\tilde S^\dagger T_{UU} \tilde
S$
to order $\tilde g^2$. Indeed, for small
reflection coefficient, the phase shift of the reflected wave is $\pi/2$
whereas in the large reflection
limit, i.e. in the DF case, the shift is $\pi$, see eq. (\ref{aa}).
It should also be pointed out that the relations between eqs. (\ref{firstT},
\ref{secondT})
 and eqs. (\ref{T1}, \ref{T2})
are valid only if $\sqrt{\sigma }$
remains much smaller than the rest mass $M$.
Indeed if this latter condition is not satisfied, one would find Doppler
effects and more
importantly, one should take into account the phases introduced by the kinetic
energy of the mirror.
This latter effect cannot be ignored when the time interval  becomes
comparable to $M/\sigma^2$. Therefore the DF results have a validity which is
limited
in both frequency ranges and time intervals. This latter condition on a maximal
interval of time
will reappear in the next section for the accelerated mirror with greater
strength owing to the relativistic
character of that trajectory.
Furthermore, it would be interesting to extend those results to higher order in
$\tilde g$. This might lead
to stronger restrictions.

Before analyzing the non-inertial model, we briefly discuss the correlations
between the
scattered momentum of the mirror and the radiation state. We just saw how these
correlations
affect the mean value of the scattered flux. To analyze the individual
correlations, we
consider the conditional value
of the flux when one knows the final position of the mirror, see eq. (\ref{w})
for the
definition of the conditional value of $T_{UU}$.
In this case the conditional flux is given
by
\be
\bk{T_{UU}^{scat}}_{\Pi_z} = { \expect {in}{\tilde S^\dagger \ \Pi_z T_{UU} \
\tilde S}{in} \over
\expect{in}{\tilde S^\dagger \ \Pi_z  \ \tilde S}{in} }
\label{w3}
 \ee
where $\Pi_z = \mbox{I}_{\phi} \ket{z}\bra{z}$ is the projector
which specifies that the final state of the
mirror is given by $\int dp e^{ipz} c^\dagger_p \ket{0}_{Mir} =\ket{z}$ and
which gives
no restriction on the radiation state.

To order $\tilde g^2$, $\bk{T_{UU}^{scat}}_{\Pi_z} $ is again given by two
terms
\ba
\bk{T_{UU}^{scat}}_{\Pi_z} &=& \tilde g^2 \vert B \vert^2
 \int_0^\infty d\om_1 \int_0^\infty d\om_2
{\sqrt{ \om_1 \om_2} \over 2\pi}e^{-i(\om_1-\om_2)(U-2z)}
 \left[ f_{\om_1}f_{\om_2}^* + g_{\om_1}g_{\om_2}^*
\right]/D
\nonumber\\
&&-\tilde g^2 \int_0^\infty d\om_1 \int_0^\infty d\om_2 {- \sqrt{ \om_1 \om_2}
\over 2\pi}
\mbox{Re} \left[
e^{-i(\om_1+\om_2)(U-2z)}
 \left(A^* B  f_{\om_1}g_{\om_2}  \right) \right]/D\ \
\label{posts}
\ea
where $D= \expect{in}{\tilde S^\dagger \ \Pi_z  \ \tilde S}{in}$ is the
probability to find the mirror at $z$
for $t=\infty$. To this order in $\tilde g$, it is evaluated without scattering
 and thus equal to $\vert \int dp e^{ipz} h_p\vert^2$. Notice that $z$ should
not be too large otherwise
one obtains $D < \tilde g^2$
which invalidates eq. (\ref{posts}).
When $D>> \tilde g^2$, one sees that the fact to have inserted the projector
$\Pi_z$ restores the
initial energy distribution, see eqs. (\ref{T1}, \ref{T2}), up to
the sign flip between the first and the second term, c.f. the discussion
above eq. (\ref{w3}).
Notice also that the final specification of the
position of the mirror introduces the phase shifts $e^{2i \om z}$, as if the
{\it initial } state of the
mirror was $\ket{z}$.


Therefore, the decoherence obtained in eq. (\ref{Tuuscat}) upon computing the
mean flux, i.e.
the vanishing of the interfering second term given in eq. (\ref{secondT}), is
"removed"
 when one knows simultaneously the final state of the mirror {\it and} the
value of the flux in this
particular channel. It is the access to these
non local correlations which reveals the "purity" of the state of the whole
system since the
momentum transfers to the mirror has mixed the states of radiation with the
mirror's state
and therefore has destroyed the
purity of the initial radiation, see eq. (\ref{khi}) and eq. (\ref{T2}).

\section{The Non Inertial Dynamical Mirror}

We shall generalize the previous results to non-inertial trajectories, in
particular
to the uniformly accelerated mirror.

Since the mirror is dynamical, the only way
to make it accelerate is to introduce an external field. The simplest case is
to consider
an electro-magnetic field coupled to the massive $\psi$ field only.
In that case, the dynamical equations (\ref{EL}) become
\ba
\left[ \partial_t^2 -\partial_z^2 \right]\phi(t,z)
&=&  \tilde g J^{\mu, A}_\psi (t, z) \ 2i\partial_\mu \phi(t, z)
\nonumber\\
\left[ (\partial_t-iA_t(t,z))^2 -(\partial_z-iA_z(t,z))^2 + M^2 \right]  \psi
(t, z) &=&
\tilde g J^\mu_\phi(t,z)\ 2(i\partial_\mu -A_\mu) \psi(t, z) \quad\ \
\label{EL2}
\ea
where the current of the $\psi $ field is now
\be
J^{\mu, A}_\psi (t, z)=\psi^\dagger(t, z) i\!\lrD_\mu
\psi(t, z)=
\psi^\dagger(t, z) \left[ i\!\lr{\partial_\mu} -2A_\mu(t,z) \right] \psi(t, z)\
\ \quad
\label{curA}
\ee
One can proceed as before and compute perturbatively the matrix elements of
$\tilde S$.
One solves the free d'Alembertian for $\phi$. For $\psi$ one solves the Klein
Gordon equation in
the presence of the $A_\mu$ field, perturbatively or non-perturbatively.
It is worthwhile to point out that the perturbative treatment, in $A_\mu$,
needs not  give the
same amplitudes for pair creation as the ones obtained non-perturbatively, see
refs.
\cite{Nik}\cite{Myrv} in that respect.

We now specialize to the case of  the constant electric field.
We proceed as follows. First we obtain the non-perturbative solutions of the
Klein Gordon
equation in the electric field. Then we compute perturbatively (in $\tilde g$)
 the scattering matix elements.
We show that to first order both in $\tilde g$ and
in the momentum transfer, these amplitudes are identical to the ones
obtained without recoil in the DF model, see eqs. (\ref{eight}). Nevertheless a
difference
in the interpretation occurs. What replaces the $\bt$ coefficient is now
directly interpretable as
the amplitude of probability to make a pair of Minkowski quanta, and no longer
as an overlap
between positive and negative frequency solutions.

We then take into account recoil effects, i.e. higher orders in the momentum
transfer.
As in ref. \cite{rec}, we shall see that the emissions of forward quanta
decohere due to the
recoils induced by these emissions. Contrary to what happens in the previous
section, the
decoherence is now inevitable owing to the properties of the spontaneously
emitted quanta.
We insist on this fact. Since the production is spontaneous and has its
properties determined
by the trajectory of the mirror, one can no longer intervene from the outset
and specify to work in
a given regime. I believe that this aspect invalidates any semi-classical
treatment of the
back reaction on the mirror trajectory.  Indeed the characteristic mean
frequency of the radiation
is precisely given in terms of this trajectory. Therefore
we are not in presence of "fast" and "slow" variables.
But such a division is implicitly assumed in
the semi-classical treatment\footnote{In ref. \cite{CV}, such a semi-classical
treatment has been analyzed.
It suffers from several difficulties, such that an acausal behavior induced by
a quantum force containing
a temporal third derivative of the mirror's trajectory. In ref. \cite{mt3}, it
was shown how this acausal behavior
disappears when the
mirror is taken to be sufficiently transparent at high frequency. In our case,
the respect of the Feyman rules
upon reflection leads to the decoherence of the emissions. This decoherence can
be viewed as a smoothing
out of the flux obtained in the absence of recoil. Therefore it eliminates the
possibility
of defining a local effective force containing a third derivative.}
and to my understanding needed for guaranteeing its
validity.

We first review the basic properties of the solutions of the Klein Gordon
equation in a
constant electric field. We work in the
homogeneous gauge: $A_t=0, A_z=Et$.
In that case, the momentum $p$ is  conserved
 and the energy $\Om(p,t)$ of  the relativistic mirror of mass $M$ is given by
\begin{equation}
\Om^2 (p, t) = M^2 + (p - Et)^2
\label{KGE}
\end{equation}
Contrariwise to the inertial case, for the accelerating mirror, the fully
relativistic equation
should be used. Indeed, it is only upon studying the "direct"
scattering, i.e. the $\al$ coefficient,
that one can choose the initial frequency high enough to build a sufficiently
localized
wave packets with respect to $1/a$. As just said, upon studying pair creation,
 the frequencies of the produced
quanta are given in terms of the classical properties of the mirror's
trajectory.
Therefore their mean frequency is precisely of the order of the acceleration.
This implies that the
minimal interval of proper time $\tau$,
during which the acceleration is constant and which is needed to have a signal
not marred by transcients
effects, should last many
$1/a$. (The same criterion is also required if one wishes to obtain a clean
Unruh effect\cite{Unr},
see \cite{MaPa}.)
The classical relativistic equations of motion of the mirror accelerating to
the left are
\begin{eqnarray}
\Om(p, t) &=& M \mbox{cosh} a \tau \nonumber\\
t - p/E &=& (1/ a)  \mbox{sinh} a \tau \nonumber\\
z - z_0  &=& -(1/ a) \mbox{cosh} a \tau
\label{eqmot}
\end{eqnarray}
The free, $\tilde g=0$, Klein Gordon equation for a mode $\psi_{p}(t,z)=
e^{ipz} \chi_{p}(t)$ is
\begin{equation}
\left[ \partial_t^2 + M^2 + (p-Et)^2 \right] \chi_{p}(t) = 0
\label{KGE2}
\end{equation}
The solutions are well known, there are cylindrical parabolic (Whittaker)
functions, see e.g. \cite{BPS}.
In what follows, we shall suppose
 that $M^2>>E$. Then the Schwinger pair production amplitude\cite{schw}
may be completely ignored since the mean density of produced pairs
scales like $e^{-\pi M^2/E}$. Furthermore,
the W.K.B. approximation for the modes  $\chi_{p}(t)$ is
valid for all $t$
and wave packets of the form
\begin{equation}
\Psi_{p}(t,z)= \int dp^\prime {e^{-(p^\prime - p)^2/2 \sigma^2} \over
( \pi \sigma^2)^{1/4}} \psi_{p^\prime}(t,z)
\label{wp}
\end{equation}
do not spread if $\sigma^2 \simeq E$ \cite{BPS}\cite{BMPPS}.
In that case, the spread  in $z$ at fixed $t$
is of the order of ${E}^{-1/2}=(M a)^{-1/2}$ for all times and thus much
smaller than the acceleration length $1/a$. Therefore these wave packets
provide, at least in the absence of scattering,
 a quantized version of a massive object following a single uniformly
accelerated trajectory.

Before analyzing the modifications of these wave packets induced by the
spontaneous production of pairs, we compute the matrix elements $\tilde S(p,
k_\om, k_\la)$.
As in eq. (\ref{tildeS}),
to first order in $\tilde g$, the amplitude to scatter an initial quantum of
momentum $k_\om$
into an outgoing one of momentum $k_\la$ is given by
 \ba
\tilde S(p, k_\om, k_\la)&=&
\delta(p'-p) \delta(k_\la-k_\om) - i \tilde g
\int_{-\infty}^{\infty} dt \int_{-\infty}^{\infty}  dz
\bra{p'} \psi^\dagger i\!\lrD_\mu \psi \ket{p} \bra{k_\la} \phi^\dagger i\!\lr
{\partial_\mu}
 \phi \ket{k_\om}
\nonumber\\
&=&
\delta(p'-p) \delta(k_\la-k_\om) - i \tilde g
\delta(p'-p -k_\om +k_\la) A(p, k_\om, k_\la)
\label{deltasss}
\ea
We still have momentum conservation owing to the homogeneous character of the
external
electric field. The amplitude $A(p, k_\om, k_\la)$ is given by
\be
A(p, k_\om, k_\la)=\int^\infty_{-\infty} dt \left[
\chi^*_{p+k_\om-k_\la}(t)
i\!\lrD_\mu \chi_p(t) \right]
(k^\mu_\om + k^\mu_\la) { e^{i(\la -\om)t} \over 4\pi \sqrt{\om \la}}
\label{overlap}
\ee
Using the W.K.B. solutions of eq. (\ref{KGE2}), $A(p, k_\om, k_\la)$ becomes
\ba
A(p, k_\om, k_\la) = \int^\infty_{-\infty} {dt \over 2 \pi}
 {e^{i\int^t dt^\prime
 \left[ \Om(p + k_\omega - k_\la, t^\prime) - \Om(p, t^\prime)
\right]}}
{ e^{i(\la -\om)t}
}\quad\quad\quad\quad\quad\quad\quad\quad\quad\quad\quad\quad&&\nonumber\\
\left[{ (\Om(p + k_\omega - k_\la, t)+\Om(p,t))(\om +\la)
-(2p + k_\omega - k_\la+2Et)(k_\om +k_\la)
\over 4 \sqrt{ \Om(p + k_\omega - k_\la,t) \Om(p, t) \om \la}} \right]&&
\label{tildB2}
\ea
compare with the inertial case, eq. (\ref{deltass}).
Using the classical equations (\ref{eqmot}), developing the phase to first
order in
$ k_\omega - k_\la$ and neglecting  the dependance in $ k_\omega- k_\la$ both
in the
numerator and in the denominator when it appears added to $Et$,
one obtains
\be
A(p, k_\om, k_\la) = \int^\infty_{-\infty} { dt \over 2 \pi} {
e^{-i(k_\om -k_\la)z_{cl}(t) + i(\la -\om)t} }
\left[ {E z_{cl}(t) (\om + \la) - (Et- p) (k_\om + k_\la) \over   2 \Om(p,t)
\sqrt{\om \la}}\right]
\quad\quad
\label{A2}
\ee
See ref. \cite{rec} for a detailed analysis of the semi-classical properties of
the W.K.B.
waves. To this order in $k_\omega - k_\la$, we emphasize that only classical
physics enters into the amplitude $A(p, k_\om, k_\la) $. Indeed, if
$\om$ is a $V$ mode and $\la$ a $U$ mode, i.e $k_\om =-\om, k_\la= \la$,
using $U=t-z_{cl}(t)$ as the dummy variable to perform the integration,
 one recovers identically (up to an irrelevant phase)
 the $\al_{\om, \la}$ coefficient of the no recoil case, eq. (\ref{K}).
Therefore the quantum
properties are encoded in higher orders in the momentum transfer.

Since the mirror is subject to the electric field, there is now spontaneous
pair creation.
The amplitude of probability to create a pair a quanta
with momenta $k_\om$ and $ k_\la$ is
\ba
\tilde S_{creat}(p, k_\om, k_\la) &=& - i \tilde g
\int_{-\infty}^{\infty} dt \int_{-\infty}^{\infty}  dz
\bra{p'} \psi^\dagger i\!\lrD_\mu \psi \ket{p} \bra{0}a_{k_\la }b_{k_\om}
 \phi^\dagger i\!\lr {\partial_\mu}
 \phi \ket{0}
\nonumber\\
&=&
 - i \tilde g
\delta(p'-p + k_\om + k_\la) B(p, k_\om, k_\la)
\label{deltass3}
\ea
where the amplitude $B(p, k_\om, k_\la)$ is given by (compare with eq.
(\ref{overlap}))
\ba
B(p, k_\om, k_\la) &=& \int^\infty_{-\infty} dt
\left[ \chi^*_{p-k_\om-k_\la}(t)
i\!\lrD_\mu \chi_p(t) \right]
(k^\mu_\la - k^\mu_\om) { e^{i(\la +\om)t} \over 4\pi \sqrt{\om \la}}
\nonumber\\
&=&A(p, - k_\om, k_\la)\\
&=&\int^\infty_{-\infty} { dt \over 2 \pi} {
e^{i(k_\om +k_\la)z_{cl}(t) + i(\la +\om)t} }
\left[ {Ez_{cl}(t) (-\om + \la) - (Et- p) (-k_\om + k_\la) \over
2 \Om(p,t) \sqrt{\om \la} }\right] \quad\quad
\nonumber
\label{B1}
\ea
where we have performed the same approximations which have lead to eq.
(\ref{A2}).
 Using again the variable $U=t-z_{cl}(t)$, we recover $\bt^*_{\om, \la}$ of eq.
(\ref{K}).
We insist on the fact that $\tilde S_{creat}(p, k_\om, k_\la)$ is interpretated
as the amplitude of probability to create the pair of quanta $b_\om$
and $a_\la$. On the contrary, $\bt_{\om, \la}$ had not such a simple
interpretation:
in terms of matrix elements
of in and out operators, it is given by
\be
\bt_{\om, \la}= \expect{0, in}{a^{\dagger out}_\la b^{\dagger in}_\om }{0, in}
\label{beta3}
\ee
It is hard to interpret the matrix element as a given pair creation act
since the frequency $\om$, the in operator and the in-vacuum are defined
in the remote past. Furthermore, if one wishes to obtain this matrix element
from the
dynamical model, it would require presumably a resummation of a certain class
of Feyman
diagrams each of which being evaluated approximatively (no loop). To determine
the class of diagrams
and the approximations used
seems a difficult task but will presumably shed light onto the approximate
nature
of the semi-classical treatment.

We have thus shown that to first order in $\tilde g$ and in the momentum
transfer
$k_\om \pm k_\la$, the matrix elements $A(p, k_\om, k_\la)$ and $B(p, k_\om,
k_\la)$
of the dynamical model coincide with the
Bogoliubov coefficients $\al_{\om, \la}$ and $\bt_{\om, \la}$ obtained in the
DF model.

We now consider higher orders in the momentum transfer.
We evaluate these higher orders using the saddle point
approximation.
The stationary phase condition of the
amplitude $A(p, k_\om, k_\la)$ indicates at which time $t^*$ one has
 conservation of the Minkowski energy:
\be
\om + \Om(p,t^*) = \la + \Om(p+k_\om-k_\la, t^*)
\label{energ}
\ee
Taking the square of eq. (\ref{energ}) and using eqs. (\ref{KGE}, \ref{eqmot}),
one obtains
\ba
\om \left[ \Om(p,t^*) + p -Et^* \right] &=& \la (\Om(p,t^*) - p + Et^* +2 \om)
\nonumber\\
\om e^{-a \tau^*} &=& \la ( e^{a \tau^*}+2 \om/M)
\label{qenerg}
\ea
This resonance condition between $\om$ and $\la$ differs from the no recoil
result,
 given by eq. (\ref{ten}) with $V_{cl}=-1/a^2U$, by the additional term on the
r.h.s.:
$2\la \om /M$. This term is quantum in character since it is linear in $\hbar$.
Its origin can be attributed to the "absorption" of the $\om$ quantum which
precedes the
emission of the $\la$ quantum. Indeed if one study, as in the accelerated
detector case, the double process
(described by a cubic interaction acting twice)
of an absorption followed by an emission,
one finds eq. (\ref{energ}) upon eliminating the energy of the intermediate
excited state\cite{rec}.

Then straightforward algebra
 gives the saddle point
approximation of $A$ and $B$.
For $\om >> \la$, i.e. in the future branch of the hyperbola, eq.
(\ref{eqmot}), one obtains
\ba
A^{s.p.}(p, \om, \la) &=&  \al^{s.p.}_{\om, \la} \ e^{i[-2\om p + \om^2]/2E}
 \nonumber\\
B^{s.p.}(p, \om, \la) &=&  \bt^{s.p. *}_{\om, \la}\  e^{i[2\om p + \om ^2]/2E}
\label{AB}
\ea
where $\al^{s.p.}_{\om, \la}, \bt^{s.p.}_{\om, \la} $ are the saddle point
expressions for the
integrals given in eqs. (\ref{K}). The additional phases are quantum in
character. It is now appropriate
to make contact
with ref. \cite{Verl} in which the quantum corrections to the Bogoliubov
coefficients evaluated at the
background field approximation are also considered. We might then say
 that the factors appearing in eqs. (\ref{AB}) arise from the "quantization of
the Bogoliubov coefficients"
here represented in their quantized version by the amplitudes $A$ and $B$.
These relations between saddle point expressions can be extended to relations
between the exact
expression of $A$ ($B$) with the exact one of $\al$ ($\bt$) given in eq.
(\ref{K}).
This is done in the Appendix. The exact
relations confirm the validity of the saddle point treatment.
What interests us is to determine the consequences of these additional phases
upon computing
the mean flux and the correlations among emitted quanta.

We first study the mean flux. We wish to compare, to order $g^2$, the flux
emitted by the
dynamical mirror with the flux
obtained in the no recoil case given in eqs. (\ref{Tuuq}, \ref{separ}). In that
case,
for the uniformly accelerated mirror,
we recall that the mean
flux vanishes identically and that the first term of eq. (\ref{separ}) is
constant in the rest frame
of the mirror.
To compute the flux in the dynamical case, it suffices to transpose what was
done in
Section 4 where we studied the scattering by an inertial mirror
to the present case of spontaneous production.

In that section, we saw that a too small $\sigma$, the inverse spread of the
mirror, easily leads
to decoherence since this effect appears
once the scattered frequency $\om $ is bigger than $\sigma$. We also mentioned
that for too large $\sigma$, one should take into account the energy of the
mirror, see discussion
after eq. (\ref{secondT}). Both restrictions appear in the present accelerated
case and play
a determinative role. Indeed, their combined effects are such that the
decoherence of successive emissions
necessarily appears before a proper time
lapse given by $\ln(M/a)/2a$ no matter what spread $\sigma$ has been chosen.
More precisely, the period of coherence is maintained during that interval only
if one deals
with minimal wave packets
characterized by $\sigma^2=E$. In order to prove this, we use the
wave packets given in eq. (\ref{wp}).
We center the mean initial momentum at $p=0$ and choose the phase of the
W.K.B solutions in
such a way that the position of the turning point of the
mean trajectory encoded in the
initial wave function is at $t=0, z_0 =0$, see eq. (\ref{eqmot}).
This choice of phases amounts to set the interaction at $t=0$ and to look only
into the
future left moving part of the hyperbola eq. (\ref{eqmot}).
 Therefore the frequency $\om$ of the $V$ mode is blue-shifted while the
frequency $\la$ of the $U$ mode in red shifted. Then only $V$ quanta will
induce
decoherence effects.

We can anticipate these effects. When a minimal wave packet is scattered
by the production of a pair in which the $V$ quantum has an energy greater than
$\sigma =E^{1/2}$,
we find, as in eq. (\ref{Tuuscat}),
that the scattered wave packet scattered no longer overlaps
with the unscattered one. Furthermore, in the rest frame of the mirror, the
mean frequencies of the quanta
spontaneously produced are of the order of
$a$. Therefore, after a proper time greather $2 a\tau= \ln (M/a)$,
the frequency of the quanta emitted forward are greather than $E^{1/2}$.
Then the successive emissions of forward quanta are completely incoherent and a
positive flux is emitted.
On the contrary, $\bk{T_{UU}}$, the flux emitted backward, is still coherent
and thus still vanishes.
This will be proven explicitly to order $\tilde g^2$; we shall see how the
recoil induced by the
blue shifted $V$-partner disappears from the expression of $\bk{T_{UU}}$.

The proof of these
facts  goes along the same lines as the development from eqs. (\ref{struct}
$\to$ \ref{secondT}).
The initial state is now $\ket{in^\prime} = \ket{Mir}\ket{0}$, see eq.
(\ref{mirror}) and eq. (\ref{wp}).
The scattered state has the following structure
\be
\tilde S \ket{in^\prime}  = \ket{in^\prime} - i \tilde g \ket{Mir', 1 pair} -
\tilde g^2 \ket{Mir'', 1 scattered\  pair}
\label{scattt}
\ee
where the linear term in $\tilde g$ is given by
\be
\ket{Mir', 1 pair} = \int dk_\om \int dk_\la \int dp \ h_{p} \ B(p, k_\om ,
k_\la)
c^\dagger_{p - k_\om -k_\la} \ket{0}_{Mir} \otimes a^\dagger_{k_\la}
b^\dagger_{k_\om} \ket{0}
\label{amplB}
\ee
see eq. (\ref{B1}) for the amplitude $B(p, k_\om , k_\la) $ to create the $\om,
\la$ pair.
We have extracted from
the terms in $\tilde g^2$ the part that shall contribute to the mean flux to
order $\tilde g^2$.
This part contains
only one created pair which has been scattered after having been created.
Neglecting once more the
complication of the time ordered product of $\tilde H_{int}$, one finds
\ba
 \ket{Mir'', 1 scattered\ pair}= \int dk_\om \int dk_\la \int d\la' \int dp \
h_{p} \ B(p, k_\om , k_\la)
A(p- k_\om -k_\la, k_\om, k_\la') &&
\nonumber\\
 c^\dagger_{p - k_\la'-k_\la} \ket{0}_{Mir} \otimes
a^\dagger_{k_\la} b^\dagger_{k_\la'} \ket{0} + ( \la \to \om )&&
\label{AmpC}
\ea
where the amplitude $A(p- k_\om -k_\la, k_\om, k_\la')$ is given in eq.
(\ref{A2}). This second scattering
has replaced $b^\dagger_{k_\om}$ by a $b^\dagger_{k_\la'}$ in the first term
and  $a^\dagger_{k_\la}$
by  $a^\dagger_{k_\la'}$ in the second one.

Then, to order $\tilde g^2$, the mean energy emitted to the left is
\ba
\langle \tilde H_V \rangle =  \bra{in^\prime} \tilde S^\dagger H_V \tilde S
\ket{in^\prime}&=&
\tilde g^2 \int dp \vert h_p \vert^2
\int_0^{\infty} d\omega \omega \int_0^{\infty} d\la \vert B ( p, \omega, \la )
\vert^2
\nonumber\\
&=& \tilde g^2
\int_0^{\infty} d\omega \omega \int_0^{\infty} d\la \vert \bt(\omega, \la )
\vert^2
\label{Hbet}
\ea
As in eq. (\ref{Hv3}), only the linear term in $\tilde g$ contributes to
$\langle \tilde H_V \rangle $.
Then the fact that $B ( p, \omega, \la )$ differs from
$\bt(\omega, \la ) $ by a phase only, see eq. (\ref{AB}),
guarantees that the mirror's wave function factorizes out
and that the total energy is unaffected by the recoils.

On the contrary the local
flux emitted to the left is affected by these phases. It is given by
\begin{equation}
\langle \tilde T_{VV} \rangle =  \bra{in^\prime} \tilde S^\dagger T_{VV} \tilde
S \ket{in^\prime} =
\langle \tilde T_{VV} \rangle_1 +
\langle \tilde T_{VV} \rangle_2
\label{T1+2}
\end{equation}
 $\langle \tilde T_{VV} \rangle_1$ is quadratic in $B$
\begin{eqnarray}
\langle \tilde T_{VV} \rangle_1 = 2 \tilde g^2 \int dp_1 \int dp_2
{e^{-(p_1)^2/2\sigma^2 }
\over
( \pi \sigma^2)^{1/4}}
{e^{-(p_2)^2/2 \sigma^2}
\over
( \pi \sigma^2)^{1/4}}
\int_0^{\infty} d\omega \int_0^{\infty} d\omega^\prime \int_0^{\infty} d\la
\int_0^{\infty}
d\la^\prime \delta( \la -\la')
&&
\nonumber\\
 \delta( p_1 + \omega - p_2 - \omega^\prime)  \left[
B ( p_1, \omega, \la )  B^*( p_2, \omega^\p, \la')
{\sqrt{\omega \omega^\prime}
\over 2\pi}
{e^{i(\omega^\prime - \omega)V}}
\right]
&&
\label{T13}
\end{eqnarray}
where the $\delta$'s of Dirac comes from momentum conservation of
the mirror and the $U$ quantum of frequency $\la$.
Similarly the interfering term, $\langle \tilde T_{VV} \rangle_2$ is
\begin{eqnarray}
\langle \tilde T_{VV} \rangle_2 =
2 \tilde g^2 \int dp_1 \int dp_2
{e^{-(p_1)^2/2\sigma^2 } \over ( \pi \sigma^2)^{1/4}}
{e^{-(p_2)^2/2 \sigma^2} \over ( \pi \sigma^2)^{1/4}}
\int_0^{\infty} d\omega \int_0^{\infty} d\omega^\prime \int_0^{\infty} d\la
\int_0^{\infty} d\la^\prime
\delta( \la -\la')&&
\nonumber\\
   \delta( p_1 + \omega + \omega^\prime -p_2)
\mbox{Re} \left[
B ( p_1, \omega, \la ) A( p_1+\om-\la, \omega^\p, \la')
{-\sqrt{\omega \omega^\prime}
\over 2\pi}
{e^{-i(\omega^\prime +\omega)V}}
\right]
&&
\label{T23}
\end{eqnarray}
where $B ( p_1, \omega, \la ) A( p_1-\om+\la, \omega^\p, \la')$ is the
amplitude
to emit two $V$ photons of frequencies $ \omega$ and $\omega^\prime$, see eq.
(\ref{AmpC}).
Notice that the
argument of the $\delta$ of Dirac arising from the overlap of the mirror states
is not the same as in eq. (\ref{T13}). It comes now from the overlap
between the twice scattered
mirror of momentum $p_1 + \omega + \omega^\prime$ with the
unperturbed one of momentum $p_2$.

Using eqs. (\ref{AB}), one can perform the integrals over $p_2$ and $p_1$. One
obtains
\ba
\langle \tilde T_{VV} \rangle_1 &=&
2 \int_0^{\infty} d\omega \int_0^{\infty} d\omega^\prime
e^{-(\omega - \omega^\prime)^2 (1/\sigma^2 + \sigma^2/E^2) /4}
\int_0^{\infty} d\la
 \left[  \bt^{s.p. *}_{\omega , \la}\bt_{\om', \la}^{s.p.}
{\sqrt{\omega \omega^\prime}
\over 2\pi}
{e^{i(\omega^\prime - \omega)V}}\right]\nonumber\\
&=&  \int_{-\infty}^{\infty}d\eta
{e^{-\eta^2/(1/\sigma^2 + \sigma^2/E^2) }\over ( \pi (1/\sigma^2 +
\sigma^2/E^2))^{1/2}}
\langle  T_{VV} (V+\eta) \rangle^{first \ term}
\label{T15}
\ea
and
\ba
\langle \tilde T_{VV} \rangle_2  &=&
- 2 \int_0^{\infty} d\omega \int_0^{\infty} d\omega^\prime
e^{-(\omega + \omega^\prime)^2 (1/\sigma^2 + \sigma^2/E^2) /4}
\int_0^{\infty} d\la
\mbox{Re} \left[ \bt^{s.p. *}_{\omega , \la}\al_{\om', \la}^{s.p.}
{\sqrt{\omega \omega^\prime}
\over 2\pi}
{e^{-i(\omega^\prime +\omega)V}}
\right]\nonumber\\
&=&  \int_{-\infty}^{\infty}d\eta
{e^{-\eta^2/(1/\sigma^2 + \sigma^2/E^2) }\over ( \pi (1/\sigma^2 +
\sigma^2/E^2))^{1/2}}
\langle  T_{VV} (V+\eta) \rangle^{second \ term}
\label{T25}
\ea
In the first equalities, the quantities in brackets are computed in
the no-recoil case in the saddle point approximation, see eq. (\ref{K}). In the
second equalities,
$\langle  T_{VV} \rangle^{first \ term}$ and $\langle  T_{VV} \rangle^{second \
term}$ are respectively
the first and second term of eq. (\ref{separ}).
Notice that the
width of the gaussian is minimal for minimal wave packets, i.e. $\sigma^2=E$,
and that the
dependence in $E$ came from the quantum phases of eqs. (\ref{AB}) only.

Thus to order $\tilde g^2$, by neglecting the time ordering upon evaluating the
scattered
state in $\tilde g^2$, and by assuming that $M>>a$, the mean flux emitted by
the dynamical
mirror can be expressed as the integral of the flux of the no-recoil case over
a region given by
$(1/\sigma^2 + \sigma^2/E^2)^{1/2}$ where $1/\sigma$ is the spread of the
initial wave packet.
Most presumably no simple relation will be found when one of the conditions
listed above is not fulfilled.
Nevertheless, the second interfering term will always be erased by the recoils
since nothing
can prevent momentum transfer nor affect the exponentially growing
Doppler shift relation between in and out frequencies.  Therefore, once $\om >
E^{1/2}$,
or $2 a\tau^*> \ln M/a$, see eq. (\ref{qenerg}), $\langle \tilde T_{VV}
\rangle_2$
 vanishes and the flux emitted
forward is now positive\footnote{The same conclusion is obtained when one
studies
the electro-magnetic flux radiated by a charged particle in a constant electric
field, see \cite{Nik}\cite{Myrv}.
This is not in contradiction with the vanishing classical reaction force in the
case of uniform acceleration. In the second
case, the trajectory is {\it a priori} given and thus it is not a solution of
the coupled Euler Lagrange equations for the
particle and the $A_\mu$ field. The important point is that one cannot use the
vanishing result of the second case
 to claim that an electron in a constant E-field does not radiate, even
locally.}
 and incoherent since the gaussian factor
$e^{-(\omega - \omega^\prime)^2 (1/\sigma^2 + \sigma^2/E^2) /4}$ appearing in $
\langle \tilde T_{VV} \rangle_1$ restricts $\om'$ to be equal to
$\om \pm (1/\sigma^2 + \sigma^2/E^2)^{-1/2}$.

Similar equations give $\bk{\tilde T_{UU}}$, the mean flux emitted to the right
by Doppler red shifted $U$ quanta.
 It is important to notice that despite the presence of hard blue shifted $V$
quanta, the fact that
$T_{UU}$
acts as the operator unity for the $V$ quanta leads to a $\delta(\om - \om')$
in the place of
the $\delta(\la- \la')$ of eq. (\ref{T23}). This delta insures that the quantum
phases
of the amplitudes $B$ and $A$ either cancel out or are reduced to low and
negligible corrections when $M>>a$.
Thus $\bk{\tilde T_{UU}}$ vanishes as in the DF model.
It would be interesting to generalize this decoupling to all order in $\tilde
g$ and to understand how generic
is this decoupling. This is because a similar decoupling might arise in the
black hole context as well.
Indeed  in that case, the out going frequencies are exponentially
blue shifted\cite{Hawk} as one approaches the Schwarzschild horizon at fixed
advanced Eddington Finkelstein $v$
 and are gravitationally coupled to the soft infalling modes.
Up to now there is no clear understanding  of whether these
interactions are negligible\cite{Wilc}, thereby legitimizing the semi classical
approximation, or
 big\cite{THooft}\cite{THooft2}\cite{Verl}\cite{EMP}\cite{Verl3}
and invalidating
the semi classical predictions.

To further investigate the consequencies of the introduction of the mirror's
dynamics,
we now consider the correlations between some specific emitted quantum and the
state of the mirror.

We start by isolating the "partner " wave function of a specific quantun.
As in eq. (\ref{Th}), we use a projector on the specific quantum.
Since the mirror is dynamical, the projector is now enlarged to
\be
\Pi_\la^{dyn} =  {\mbox{ I}}_{Mir} \otimes  {\mbox{ I}}_{b} \otimes
\int_0^\infty \! d\la\
f_\la a_\la^{\dagger} \ket{0_a} \bra{0_a}
\int_0^\infty \! d\la^\p\
f_{\la^\p}^* a_{\la^\p}
\label{Th2}\ee
where ${\mbox{ I}}_{Mir} $ is the operator unity on the mirror's Fock space.
To order $\tilde g^2$, the probability to find such a state is given by $\tilde
g^2 P_{\Pi_\la}$, see
eq. (\ref{Pi}), by virtue
of eqs. (\ref{AB}) which relate the dynamical matrix elements to the former
Bogoliubov coefficients.
Again one sees that the fact that the probability is a sum of squares leads to
almost no
modification even when $\la$ describes a $V$ quantum whose energy is much
bigger than $M$. This is because
the corrections of the norm of the amplitudes
are given by terms in $\la^n \partial_\la^n \ln \Om(p -\la , t)$. These
 remain of the order of $a/M$ for all times.

Nevertheless, the "partner" of the quantum  $\ket{\bar \la }=
\int d\la f_\la a_\la^{\dagger} \ket{0_a}$ is no
longer given by eq. (\ref{partner}) but by
\ba
\ket{partner,\ \bar \la}^{dyn} &=& \bra{0_a} \int d\la f^*_\la a_\la \ \tilde S
\ket{0} \ket{Mir}
\nonumber\\
&=& \int_{-\infty}^\infty dp h_{p} \int_0^\infty d\la  f_\la^* \int_0^\infty
d\om B(p,\om, \la) c^\dagger_{p+\om-\la} \ket{0}_{Mir} \otimes  b_\om^\dagger
\ket{0_b}
\label{partnerdyn}
\ea
see eq. (\ref{amplB}).
We see that we must have
both $\om$ and $\la$ are much smaller than $\sigma$, the width in $p$ encoded
in $h_p$,
 in order to be allowed to factorize out the initial mirror's wave function
$\ket{Mir}$, eq. (\ref{mirror}),
without enjuring the result.
In that case, one recovers exactly the partner as defined in eq.
(\ref{partner}).
On the contrary, when $\om- \la$ is comparable to $\sigma$,
we can no longer factorize out the mirror wave function and the "partner" is
truly a combined (entangled) state of the anti particle and the mirror's state.
This is precisely what leads
to decoherence and what happened
for the uniformly accelerated mirror since the forward frequencies are blue
shifted according to
$\om = a e^{a\tau^*}$, see eq. (\ref{qenerg}).

Another way to investigate the correlations is to consider the conditional
value of $T_{\mu \nu}$
correlated to a particular final state. As stressed in \cite{MaPa}, the matrix
elements $\bk{ T_{\mu \nu}}_{\Pi} $ given in eq. (\ref{w}) change in time from
an off-diagonal
matrix element when one inquires into the value of $ T_{\mu \nu}$ before the
interaction
occurs to a more usual expectation value when one evaluates the conditional
value
after the interaction has occurred. Therefore, by virtue of the fact that the
additional phases
introduced by the recoils play an important role only in interfering
situations, the conditional
value of $T_{\mu \nu}$ is completely washed out in the past once
the frequency $\om$ or $\la$ is greather than $\sigma$.
On the contrary, in the future, the conditional value stays
almost unaffected
by the recoil of the mirror. The easiest way to understand this double
behavior, is to compare
the momentum transfer to the mirror in  both situations.

We consider the case in which one selects a $U$ quantum by the
projector $\Pi_\la^{dyn}$, see eq. (\ref{Th2}), and we evaluate the conditional
value of $T_{VV}$
to order $\tilde g^2$, see eq. (\ref{w3}). We first analyze the structure of
 $\bk{ T_{VV}}_{\Pi} $ before the interaction occurs, i.e.
on the right of the mirror. We
obtain that the interacting hamiltonian acts linearly on both sides of the
projector  $\Pi_\la^{dyn}$.
On its right, one finds a matrix element of this type: $  \bra{Mir'}\bra{\bar
\la}\bra{\om}
 \tilde S \  T_{VV}\ket{0}\ket{Mir}$
since one starts from $\ket{0}\ket{Mir}$ and select the state $\ket{\bar \la}$.
The state $\bra{\om}$ represent the
unscattered $V$ quantum created by $T_{VV}$. Therefore the momentum of the
mirror in the
bra $\bra{Mir'}$ is $p-\om'-\bar \la$ where $\om'$ is the other $V$ quantum
created by $T_{VV}$ acting
on Minkowski vacuum. This latter gets
reflected with amplitude $A(p, \om', \bar \la)$, eq. (\ref{deltasss}),
to constitute the selected $U$ quantum $\ket{\bar \la}$.
 On the left of the projector $\Pi_\la^{dyn}$, the matrix element is of the
form
 $\bra{Mir}\bra{0} \tilde S \ket{\bar \la }\ket{\om}\ket{Mir''}=B(p, \om,\bar
\la)$.
In the ket $\ket{Mir''}$, the momentum of the mirror
is $p+\om -\bar \la$ since both $\om$ and $\bar \la$ are created.
Therefore the overlap between the two states of the mirror vanishes once the
$V$ quantum
$\om+\om'$ is greather than $\sigma$ exactly as for the $\bk{T_{VV}}_2$ term in
eqs. (\ref{secondT}, \ref{T25}).

We now consider $\bk{ T_{VV}}_{\Pi} $ after the creation act, on the
left of the mirror. As before we find that the interacting hamiltonian acts
linearly on both sides of the projector.
But the matrix element on the right of the projector has now the following
structure:
$\bra{Mir'}\bra{\om}\bra{\bar \la}
T_{VV} \tilde S  \ket{0}\ket{Mir}$. This matrix element
is governed by the amplitude $B(p, \om' , \bar \la) $ to create the pair $\bar
\la, \om'$ and the operator
$T_{VV}$ replaces the quantum $\om'$ by $\om$.
 On the left hand side of the projector nothing changes. Thus the mirror's
momentum is now  $p+\om' -\bar \la$
in the bra $\bra{Mir'}$ and still $p+\om -\bar \la$ in the ket $\ket{Mir}$.
Then the overlap between the two states of the mirror
constraints $\om -\om'$. This is exactly what
happened for $\bk{T_{VV}}_1$
in eq. (\ref{T15}).

Thus we have proven that the conditional value behaves, in the past, like an
interfering
"washable" term in which the sum of the blue shifted frequencies appear, and,
in the future, like the first term in the decomposition of the mean in which
the difference
of the momenta occur. It is also important to realize that the situation for
the conditional value of $T_{UU}$
is not the same  even though
one obtains expressions with exactly the same structure
under the replacement of the $V$ momentum $\om$ by the $U$ momentum $\la$.
For $\bk{ T_{UU}}_{\Pi} $, the momentum transfer
that made $\bk{ T_{VV}}_{\Pi} $
to vanish in the past, i.e. $2\om$, is now replaced by $2\la$ which is a very
low Doppler
red shifted frequency. Therefore, the conditional value of $T_{UU}$ is
unchanged
(recall that one considers
only the future part of the hyperbola, for $t>t_{t.p.}$, see eq.
({\ref{eqmot})).
For the same reasons, we found that the mean flux in the $U$ direction was
unchanged,
 see discussion in the paragraph before eq. (\ref{Th2}).


\vskip 1. truecm

{\it Conclusions}

\noindent
In this article we showed how decoherence effects occur when one takes into
account
the dynamics of the scattered agent and how they modify the properties of the
matrix elements
of  $T_{\mu \nu}$.
In particular, when the frequency of the spontaneously produced quanta grows
due to some
Doppler effect, these effects inevitably lead to an incoherent
positive flux as well as to vanishing conditional values of the flux in the
past of the creation act.
We believe that both consequences are generic in the sense that they will also
occur in
 cosmological contexts and in the black hole evaporation process. Indeed in
both cases, in the background field
approximation, when one traces
backwards a quantum, its frequency is blue shifted according to the
gravitational properties of the background.
This classical relation will certainly be modified once the momentum energy of
the quantum will dominate
the background mean energy. Then the local correlations between the presence of
this late quantum and
the field configurations at early times which are found when one uses the
background field approximation
will be equally washed out by the recoil of the geometry treated dynamically.

The crucial point which remains to be done is to determine the consequences of
the modifications induced by these
recoil effects. Indeed, to the lowest order in $\tilde g$,  we saw that the
probabilities to find a specific event
as well as the expectation values expressible as sum of squares of amplitutes
are almost unaffected by the
recoil effects. Therefore, one must determine if the higher orders in $\tilde
g$ will sufficiently
modify the dynamics at later times that these expectation values will no longer
be correctly approximated
 by their lowest order expressions.

\vskip 1. truecm
{\bf Acknowlegdments    }
I would like to thank M. T. Jaekel for explaining me his work on partially
reflecting mirrors.
I would also like to thank Cl. Bouchiat for numerous clarifying conversations.
Finally I thank R. Brout, S. Massar and Ph. Spindel for very useful
discussions.

\section{Appendix: The Exact Amplitudes}

The goal of this Appendix is to confirm the validity of eqs. (\ref{AB}) which
relate the saddle point
expressions for the scattering amplitudes $A$ ($B$) to the saddle point
expressions for the Bogoliubov coefficients
$\al$ ($\bt$) by finding a similar relations between exact expressions for $A$
and for $\al$.

To this end we shall use the integral representations
of the exact solutions of eq. (\ref{KGE2}) rather than the W.K.B.
 approximation which were used in eq. (\ref{A2}).
The integral representation of an in mode with asymptotic unit initial current
coming from $z=-\infty$
is given by
\be
\chi^{in}_{p}(t)=N \int_{-\infty}^0 du (-u)^{{iM^2 \over 2E}-{1 \over 2}}
 e^{-iE \left[ u^2/4 + (t-p/E)u +  (t-p/E)^2 /2\right]}
\label{App1}
\ee
where $N^{-1}$ is given by $e^{\pi M^2/E} \Gamma(1/2 + iM^2 /2E)$, see
\cite{PB1}.
The dummy variable $u$ is classically related to the time $t$ and the energy
$\Om(p,t)$, eq. (\ref{KGE}), by
$u= t -Ep -\Om(P, t) $.  $u /\sqrt{2E}$ plays for this inverse harmonic
potential case a role very similar
to the annihilation operator $a=(p + iq)/\sqrt{2}$ of the harmonic oscillator.
Similarly, the integral representation of an out mode with asymptotic unit
final current directed towards $z=-\infty$ is
\be
\label{App2}
\chi^{out}_{p}(t)=N \int_{-\infty}^0 du (-u)^{-{iM^2 \over 2E}-{1 \over 2}}
 e^{iE \left[ u^2/4  - (t-p/E)u +  (t-p/E)^2/2 \right]   }
\ee
see \cite{PB1} for more details.

Therefore the amplitude $A(p, \om, \la)$, see eq. (\ref{overlap}), is given by
\ba
A(p, \om, \la) &=& N^2 \int_{-\infty}^{\infty} {d\tilde t  \over 4\pi}
\int_{-\infty}^0 du_1  \int_{-\infty}^0 du_2
{e^{-i(\om -\la) \tilde t} \over \sqrt{\om \la}} e^{-i(\om -\la) p/E
}(-u_1)^{{iM^2 \over 2E}-{1 \over 2}}
(-u_2)^{{iM^2 \over 2E}-{1 \over 2}}
\nonumber\\
&&\ \quad (8 \la i\partial_{u_1} e^{-iE \left[ u_1^2/4 + \tilde t u_1 + \tilde
t^2/2 \right]})
 e^{-iE \left[ u_2^2/4  - (\tilde t-\om/E-\la/E) u_2 +  (\tilde
t-\om/E-\la/E)^2/2 \right]}
\label{App3}
\ea
where we have defined $\tilde t = t- p/E$. The derivative factor $ 8\la
i\partial_u$
is a useful reexpression of the matrix
elements of the current operators $J^\mu_\phi J_\mu^{A, \psi}$.
Performing the gaussian integration over $\tilde t $, one gets
\ba
A(p, \om, \la) &=& {N^2 \over \sqrt{2 \pi E }}
{e^{-i(\om -\la) p/E}
 \over \sqrt{\om \la}}  \int_{-\infty}^0 du_1  \int_{-\infty}^0 du_2
(-u_1)^{{iM^2 \over 2E}-{1 \over 2}}
(-u_2)^{{iM^2 \over 2E}-{1 \over 2}}
\nonumber\\
&& \quad\quad ( 4 \la i\partial_{u_1}
e^{-iE \left[ u_1 u_2/2 + u_1 \om/E + u_2 \la/E - \om^2 /E^2 + (\om
-\la)^2/2E^2 \right]})
\label{App4}
\ea
Introducing the variable $\delta = (E/M)^2 u_1 u_2= a^2 u_1 u_2$ and
integrating par part, one has
\ba
A(p, \om, \la) &=&  e^{-i(\om -\la) p/E} e^{i \om^2 /E -i (\om -\la)^2/2E }
{N^2 \over \sqrt{2 \pi E }}
 (M-ia)  \int^{\infty}_0 d\delta \ \delta^{{iM^2 \over 2E}-{3 \over 2}}
e^{-{iM^2 \delta \over 2E}}
\nonumber\\
&& \quad\quad \quad\quad
\int_{-\infty}^0 du_2 \ e^{-i\la u_2} e^{-i\om \delta /a^2u_2} {2\la \over
\sqrt{\om \la}}
\nonumber \\
&=&
 e^{-i(\om -\la) p/E } e^{i \om^2 /E - i(\om -\la)^2/2E } {2 \pi N^2 \over
\sqrt{2 \pi E }} ({M}-{ia})
\int^{\infty}_0 d \delta  \ \delta^{{iM^2 \over 2E}-{1}} e^{-{iM^2 \delta \over
2E}}
\nonumber \\ &&
\quad\quad \quad\quad ({1 \over a \pi})
K_1\left[ 2i ({\om \la \delta \over a^2})^{1/2} \right]
\label{App5}
\ea
see eq. (\ref{K}) for the definition of the Bessel function.
In the limit $M^2/E \to \infty$ at fixed $a =E/M$ (i.e. $\hbar \to 0$) and for
${\om \la / a^2}= {\cal O}(1)$,
 the integral over $\delta$ gets its contribution
from the saddle region $\delta =1 \pm (a/M)^{1/2}$ and the normalisation factor
$N \to 1/\sqrt{2 \pi}$.
In that case, one finds
\be
A(p, \om, \la) =e^{-i(\om -\la) p/E} e^{i (\om^2 -\la^2)/2E} \al(\om, \la)
\label{App6}
\ee
Therefore, when $\om >> \la$, one recovers eq. (\ref{AB}).
It is interesting to notice that the dummy variable $u_2$ is classically equal
to mirror's coordinate $U=-e^{a \tau}/a$
and plays exactly the role of this classical variable
in eq. (\ref{App5}). On the contrary the variable $\delta = a^2 u_1 u_2$ is
classically
equal to $1$ (since $u_1$ is classically equal to $-V_{cl}(U) = 1/a^2 U$)
and acts as a quantum spread around the classical trajectory. This spread is
controlled by $(M/\hbar a)^{-1/2}$
and has nothing to do
with the spread in position of the initial wave packet, rather it is here
induced by the specification of $\om$ and $\la$
as well as the conservation of momentum and energy (forced by the  integral
over $\tilde t$).


\begin{thebibliography}{999}


\bibitem{Hawk} S. W. Hawking, Commun. Math. Phys. {\bf  43} (1975) 199

\bibitem{new} K. Freedenhagen and R. Haag, Commun. Math. Phys. {\bf 127} (1990)
273

\bibitem{THooft} G. 't Hooft, Nucl. Phys. B {\bf 256} (1985) 727

\bibitem{Jacobson} T. Jacobson, Phys. Rev. D {\bf 44} (1991) 1731,
Phys. Rev. D {\bf 48} (1993) 728

\bibitem{Waldbook}  R. M. Wald {\it Quantum Field Theory in Curved
Spacetime and Black Hole Thermodynamics} University of Chicago Press (1994)

\bibitem{Bardeen} J. M. Bardeen, Phys. Rev. Lett. {\bf 46} (1981) 382

\bibitem{York} J. W. York, in {\it{Quantum Theory of Gravity}} ed. by
S. Christensen, Adam Hilger, Ltd., Bristol (1984)

\bibitem{Massar2}  S. Massar, {\em The semi classical back reaction to black
hole evaporation} $\quad \quad \quad$ preprint ULB-TH 94/19, gr-qc/9411039

\bibitem{PP} R. Parentani and T. Piran, Phys. Rev. Lett. {\bf 73} (1994) 2805

\bibitem{DFU}Davies P. C. W., Fulling S. and  Unruh W.G., Phys. Rev. D {\bf 13}
(1976)
2720

\bibitem{THooft2}C. R. Stephens, G. 't Hooft and B. F. Whiting,
Class. Quant. Grav. {\bf 11} (1994) 621

\bibitem{EMP} F. Englert, S. Massar and R. Parentani,
Class. Quantum Grav. {\bf 11} (1994) 2919

\bibitem{MaPa} S. Massar and R. Parentani {\em From Vacuum Fluctuations to
Radiation: Accelerated Detectors and Black Holes. (2)}, preprint ULB-TH 94/02,
LPTENS 95/07 (1995) gr-qc/9502024

\bibitem{Verl3} Y. Kiem, H. Verlinde, E. Verlinde {\it Quantum Horizons and
Complementarity.}
CERN-TH-7469-94, hep-th/9502074

\bibitem{DF} P. C. W. Davies and  S. A. Fulling, Proc. R. Soc. London {\bf A
356} (1977) 237

\bibitem{rec}R. Parentani, {\em The Recoils of the Accelerated Atom
and the Decoherence of its Fluxes}, preprint (1995) LPTENS 95/02,
gr-qc 9502030, {\em to be published in Nucl. Phys. B.}

\bibitem{CV} T. D. Chung and H. Verlinde, Nucl. Phys. B {\bf 418} (1994) 305

\bibitem{BD}N.D. Birrel and P.C.W. Davies, {\it Quantum Fields in Curved
Space},
Cambridge University Press (1982).

\bibitem{Grov} P. Grove, Class. Quant. Grav. {\bf  3} (1986) 793

\bibitem{Unr}  W. G. Unruh, Phys. Rev. D {\bf  14} (1976) 870

\bibitem{UnrW} W. G. Unruh and R. M. Wald, Phys. Rev. D {\bf 29} (1984) 1047

\bibitem{grove} P. Grove, Class. Quant. Grav. {\bf  3} (1986) 801

\bibitem{Unruh92} W. G. Unruh, Phys. Rev. D {\bf 46} (1992) 3271

\bibitem{AM} J. Audretsch and R. M\"uller,  Phys. Rev. D {\bf 49} (1994)
 4056; Phys. Rev. D {\bf 49} (1994) 6566;
Phys. Rev A {\bf 50} (1994) 1755

\bibitem{mt}  M. T. Jaekel and S. Reynaud, Phys. Lett. A {\bf 180} (1993) 9

\bibitem{MTW} Misner C.W., Thorne K.S. and Wheeler J.A., {\em Gravitation},
Freeman,
San Francisco (1973)

\bibitem{ker} B. S. de Witt Phys. Rep.  {\bf 19 C} (1975) 297

\bibitem{GO} R. Brout, S. Massar, R. Parentani and Ph. Spindel, {\em A
Primer for Black Hole Quantum Physics} ULB-TH 95/02, ULM-MG 95/01,
LPTENS 95/03  (1995) Phys. Rep. {\em in the press.}

\bibitem{BMPPS} R. Brout, S. Massar, R. Parentani, S. Popescu and
Ph. Spindel,  Phys. Rev. D {\bf 52} (1995) 1119

\bibitem{Boul}Boulware D. G., Annals of Physics {\bf 124} (1980) 169

\bibitem{Full} S. A. Fulling, Phys. Rev. D {\bf  7} (1973) 2850

\bibitem{Strom} A. Strominger, {\it Les Houches Lectures on Black Holes}
(1994), hep-th/9501071

\bibitem{Wil} F. Wilczek, {\it Quantum Purity at Small Price: Easing a Black
Hole Paradox}
IASSNS-HEP-93, hep-th/9302096

\bibitem{Carl} R. Carlitz and R. Wiley, Phys. Rev. D {\bf 36} (1987) 2327

\bibitem{RSG} D. Raine, D. Sciama and P. Grove, Proc. R. Soc. A {\bf 435}
(1991)
205

\bibitem{MPB} S. Massar, R. Parentani and R. Brout,
Class. Quant. Grav. {\bf 10 } (1993) 385

\bibitem{mt2}  M. T. Jaekel and S. Reynaud, J. Phys. I  France {\bf 167} (1992)
1

\bibitem{Nik} A. I. Nikishov, Sov. Phys. JETP {\bf 32} (1971) 690

\bibitem{Myrv} N. P. Myhrvold, Ann. of Phys. {\bf 160} (1985) 102

\bibitem{mt3}  M. T. Jaekel and S. Reynaud, Phys. Lett. A {\bf 167} (1992) 227

\bibitem{BPS} R. Brout, R. Parentani and Ph. Spindel, Nucl. Phys. {\bf B353 }
(1991) 209

\bibitem{schw} J. Schwinger, Phys. Rev. {\bf 82} (1951) 664

\bibitem{Verl} K. Schoutens, H. Verlinde,
 E. Verlinde, ``{\it Black Hole Evaporation and Quantum Gravity}"
Preprint CERN-TH.7142/94, PUPT-1441, (1994), hep-th/ 9401081.

\bibitem{Wilc} P. Kraus and F. Wilczek,
Nucl. Phys. B433 (1995) 403

\bibitem{PB1} R. Parentani and R. Brout, Nucl. Phys. B388 (1992) 474

\end{thebibliography}
\end{document}